\documentclass[%
 reprint,
 amsmath,amssymb,
 aps,
 pre,
 floatfix,
]{revtex4-2}
\usepackage{nicematrix}
\usepackage{mathtools}
\usepackage{graphicx}
\usepackage{dcolumn}
\usepackage{xcolor}
\usepackage{bm}
\usepackage[utf8]{inputenc} 
\usepackage{url}
\usepackage{amsmath}
\usepackage{amssymb}    
\usepackage{graphicx}
\usepackage{listings}
\usepackage{hyperref}

\newcommand{\cs}{c_s}

\newenvironment{eqnal}{\equation\aligned}{\endaligned\endequation}

\newcommand{\fortranl}[1]{\lstinline[language={[90]Fortran}]{#1}}

\begin{document}

\preprint{APS/123-QED}

\title{Deep modelling of plasma and neutral fluctuations from gas puff turbulence imaging}

\author{A. Mathews$^1$} \email{mathewsa@mit.edu} \author{J.L. Terry$^1$} \author{S.G. Baek$^1$} \author{J.W. Hughes$^1$} \author{A.Q. Kuang$^1$} \author{\\B. LaBombard$^1$} \author{M.A. Miller$^1$} \author{D. Stotler$^2$} \author{D. Reiter$^3$} \author{W. Zholobenko $^4$} \author{M. Goto$^5$}
\affiliation{$^1$Plasma Science and Fusion Center, Massachusetts Institute of Technology, Cambridge, Massachusetts 02139, USA
}%
\affiliation{$^2$Princeton Plasma Physics Laboratory, Princeton, New Jersey 08540, USA}
\affiliation{$^3$Institut f{\"u}r Laser- und Plasmaphysik, Heinrich-Heine-Universit{\"a}t, D{\"u}sseldorf 40225, DEU}
\affiliation{$^4$Max-Planck-Institut f{\"u}r Plasmaphysik, Garching 85748, DEU}
\affiliation{$^5$National Institute for Fusion Science, Toki 509-5292, JAP}

\date{\today}

\begin{abstract}

The role of turbulence in setting boundary plasma conditions is presently a key uncertainty in projecting to fusion energy reactors. To robustly diagnose edge turbulence, we develop and demonstrate a technique to translate brightness measurements of HeI line radiation into local plasma fluctuations via a novel integrated deep learning framework that combines neutral transport physics and collisional radiative theory for the $3^3 D - 2^3 P$ transition in atomic helium. The tenets for experimental validity are reviewed, illustrating that this turbulence analysis for ionized gases is transferable to both magnetized and unmagnetized environments with arbitrary geometries. Based upon fast camera data on the Alcator C-Mod tokamak, we present the first 2-dimensional time-dependent experimental measurements of the turbulent electron density, electron temperature, and neutral density revealing shadowing effects in a fusion plasma using a single spectral line.

\end{abstract}

\maketitle


\section{\label{sec:level1}Introduction}

Diagnosing edge plasmas is an essential task towards testing turbulence models and better understanding plasma fueling and confinement in fusion devices. Gas puff imaging (GPI) of turbulence is a widely applied experimental diagnostic that captures line emission based upon the interaction of neutrals with the hot plasma. As a technique with decades of application in a range of settings \cite{Zweben_2017}, optical imaging of fluctuations provides a view of turbulent plasma transport. This transport is critical to the operation and energy gain of nuclear fusion reactors, but interpretation (e.g. velocimetry \cite{velocimetry_GPI}) of these fluctuations to directly test reduced physics models is not always straightforward. By tracing the atomic theory underlying the nonlinear dynamics of observed line emission, we outline a novel spectroscopic method for understanding turbulent fluctuations based upon high-resolution visible imaging of plasma-neutral interactions.

The plasma edge in magnetic fusion devices is characterized by neighbouring regions: confined plasmas where temperatures can exceed 10$^{6}\text{ K}$, and the colder scrape-off layer (SOL) where gaseous particles may not be completely ionized. These regions exist tightly coupled to one another and inseparable in many respects. Consequently, accounting for neutral transport in conjunction with ion and electron turbulence is essential in wholly analyzing boundary plasma fluctuations. Edge turbulence is characterized by a broadband spectrum with perturbation amplitudes of order unity and frequencies ranging up to 1 MHz. Edge localized modes and intermittent coherent structures convecting across open field lines can be responsible for significant particle losses and plasma-wall interactions that strongly influence operations. To model the vast dynamical scales present in fusion plasmas accordingly requires sufficiently complex modelling techniques. In this work, we introduce custom neural architectures within a multi-network deep learning framework that bounds outputs to abide by collisional radiative theory \cite{Fujimoto1979ACM,GOTO_2003} and searches for solutions consistent with continuity constraints on neutral transport \cite{Wersal_2017}. Learning nonlinear physics via optimization in this way provides a new way to examine edge turbulence using experimental data from GPI. While our methodology is not fixed to any device, this paper focuses on 2-dimensional experimental brightness measurements from open flux surfaces on the Alcator C-Mod tokamak \cite{Hutch_CMod, Alcator_Greenwald}, where we find a good signal-to-noise ratio of localized light emission. Recent advancements in validation techniques of reduced turbulence theories \cite{Mathews2021PRE,Mathews2021PoP} emphasize the importance of comprehensive diagnostic coverage of electron dynamics on turbulent spatial and temporal scales. To this end, we describe the first calculations of the 2-dimensional turbulent electron density, electron temperature, and neutral density fields that self-consistently include fluctuation-induced ionization using only observations of the 587.6 nm line via fast camera imaging. With several possible extensions to the deep learning framework identified, our experimental analysis technique paves new ways in systematically diagnosing edge plasma turbulence.

To demonstrate this framework, we evaluate the validity of collisional radiative theory in conditions relevant to fusion plasmas for atomic helium line emission in Section \ref{sec:level2}, overview the experimental setup for GPI on the Alcator C-Mod tokamak in \ref{sec:level3}, outline a custom physics-informed machine learning optimization technique designed for turbulence imaging in Section \ref{sec:level4}, present results from the analysis applied to experimental fast camera data in section \ref{sec:level5}, and conclude with a summary and future outlook in Section \ref{sec:level6}.

\section{\label{sec:level2}Time-dependent analysis of quantum states in atomic helium}

The electronic transition from $3^3 D$ to $2^3 P$ quantum states in atomic helium results in photon emission with a rest frame wavelength of 587.6 nm. Atomic physics modelling of this line radiation in a plasma correspondingly requires tracking all relevant electron transition pathways that can populate or depopulate $3^3 D$. Our starting point in this analysis is to consider the full rate equations where the evolution of a quantum state $p$ follows

\begin{figure}[ht]
\includegraphics[width=1.0\linewidth]{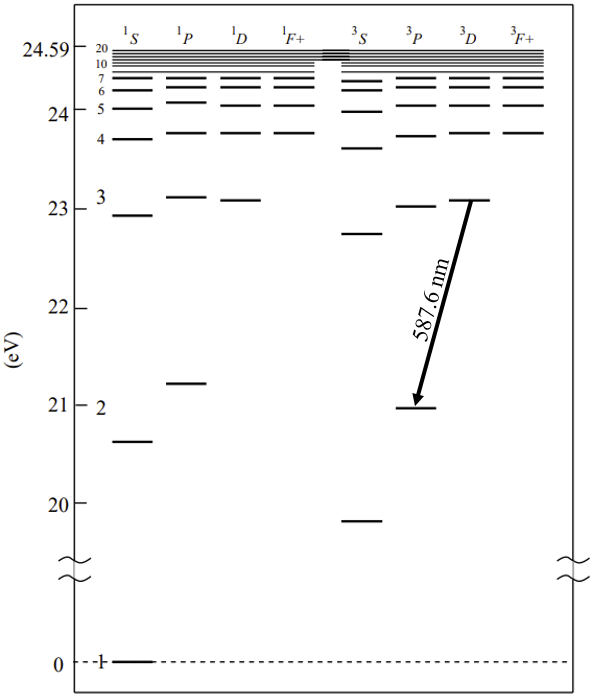}
\caption{\label{HeI_states}Energy level diagram for atomic helium considered in the calculations. An arrow connects $3^3D \rightarrow 2^3P$, which is the origin of the 587.6 nm photon emission. The labels $^{1,3}F+$ denote the quantum states representing all levels with $L \geq 3$. Figure reprinted from \cite{GOTO_2003} with permission from Elsevier.}
\end{figure}

\begin{eqnal}\label{eq:full_rate_eqn}
\frac{dn(p)}{dt} &= \sum\limits_{q \neq p} \lbrace {C(q,p)n_e + A(q,p)} \rbrace n(q) \\&- \lbrace {\sum\limits_{q \neq p} C(p,q)n_e + \sum\limits_{q < p} A(p,q)} + {S(p)n_e }\rbrace n(p) \\&+ \lbrace {\alpha(p)n_e + \beta(p) + \beta_d(p)} \rbrace n_i n_e,
\end{eqnal}

where $n(p)$ is the population density of the $p = n^{2S + 1} L$ state, in which $n$ is the principal quantum number, $S$ is the spin, and $L$ is the orbital angular momentum quantum number. Similarly, $q$ is another quantum state with the notation $q < p$ indicating that $q$ lies energetically below $p$. Eq. \eqref{eq:full_rate_eqn} includes the spontaneous transition probability from $p$ to $q$ given by the Einstein A coefficient $A(p,q)$, electron impact transitions $C(p,q)$, electron impact ionization $S(p)$, three-body recombination $\alpha(p)$, radiative recombination $\beta(p)$, and dielectronic recombination $\beta_q(p)$, with $n_e$ and $n_i$ denoting the electron density and hydrogen-like He$^+$ density, respectively. All aforementioned rate coefficients except $A(p,q)$ have a dependence on the electron temperature ($T_e$) that arises from averaging cross-sections over a Maxwellian velocity distribution for electrons, which are based upon calculations with the convergent close-coupling (CCC) \cite{CCC1,CCC2,CCC3} and R-matrix with pseudostates (RMPS) \cite{RMPS} methods using high precision calculations of helium wavefunctions \cite{Drake1,Drake2}. For application in a numerical framework, we follow \cite{GOTO_2003,Zholobenko_2021} to model atomic helium with a corresponding energy level diagram visualized in Figure \ref{HeI_states}. All quantum states with $L \leq 2$ are resolved for $n < 8$ while states with $L \geq 3$ are bundled together into a single level denoted as ``$F+$". For $n \geq 8$, $L$ is not resolved, while those with $n \geq 11$ are approximated as hydrogenic levels with statistical weights twice those of hydrogen. Quantum states up to $n=26$ are included with $n \geq 21$ being given by the Saha-Boltzmann equilibrium \cite{McWhirter_1963,Fujimoto1979ACM,Zholobenko_2021}. For application in magnetized plasmas (e.g. tokamaks), where rate coefficients vary with magnetic field strength due to wavefunction mixing, spin-orbit interactions are included to account for mixing between the singlet and triplet fine structure levels \cite{GOTO_2003,Zholobenko_2021}. Finite fields largely influence the modelling of metastable species and higher orbital quantum numbers. To quantify radiation trapping effects, the dimensionless optical depth for a Doppler-broadened line transition between states $j \rightarrow k$ can be expressed as \cite{Huba2013} 

\begin{eqnal}\label{eq:optical_depth}
\tau_{j \rightarrow k} = 5.4 \times 10^{-3} f_{j \rightarrow k} \lambda_{j \rightarrow k} n_j (\mu_j/T_j)^{\frac{1}{2}} L
\end{eqnal}

where $f_{j \rightarrow k}$ is the absorption oscillator strength, $\lambda_{jk} \ [\text{nm}]$ is the line center wavelength, $\mu_j$ is the mass ratio of the emitting species relative to a proton, $L \ [\text{cm}]$ is the physical depth of the gas along the viewing chord, and $n_j \ [10^{13} \ \text{cm}^{-3}]$ and $T_j \ [\text{eV}]$ are the density and temperature, respectively, of particles in state $j$. For 587.6 nm line emission with a thermal helium puff in conditions relevant to magnetic confinement fusion devices, where $f_{2^3P \rightarrow 3^3D} \sim 0.6$ \cite{HeI_coeff}, $\tau_{2^3P \rightarrow 3^3D} \ll 1$ resulting in the plasma edge region being optically thin for our spectroscopic analysis of a localized gas puff \cite{Zholobenko_2021,Zweben_2017}.

The rate equations \eqref{eq:full_rate_eqn} for an optically thin plasma can be equivalently expressed in matrix form as \cite{STOTLER_2007,Zholobenko_2021}

\begin{eqnal}\label{eq:matrix_full_rate_eqn}
\frac{d{\bf{n}}}{dt} = {\bf{M}}(n_e,T_e) {\bf{n}} + {\bf{\Gamma}}(n_e,T_e,n_i)
\end{eqnal}

where ${\bf{n}}$ is a vector of the $N$ atomic states, ${\bf{M}}$ represents the $N \times N$ matrix of rates for collisional ionization, excitation, de-excitation, radiative decay, and recombination as above, and ${\bf{\Gamma}}$ symbolizes sources. Since time-evolving every state in atomic helium is computationally expensive, effective atomic physics models known as collisional radiative (CR) theories are often constructed. This involves separating the $N$ states into $P$ and $Q$ spaces of sizes $N_P$ and $N_Q$, respectively, such that \eqref{eq:matrix_full_rate_eqn} becomes

\begin{eqnal}\label{eq:matrix_P_Q}
\frac{d}{dt}
\begin{bmatrix}
   {\bf n_{P}} \\
   {\bf n_{Q}}
\end{bmatrix}
=
\begin{bmatrix}
   {\bf M_{P}} & {\bf M_{PQ}} \\
   {\bf M_{QP}} & {\bf M_{Q}}
\end{bmatrix}
\begin{bmatrix}
   {\bf n_{P}} \\
   {\bf n_{Q}}
\end{bmatrix}
+
\begin{bmatrix}
   {\bf \Gamma_{P}} \\
   {\bf \Gamma_{Q}}
\end{bmatrix}
= 
\begin{bmatrix}
   {\frac{d\bf n_{P}}{dt}} \\
   {0}
\end{bmatrix}
\end{eqnal}

By taking the $Q$ space to be time-independent, under the expectation that they evolve on timescales faster than those of plasma turbulence fluctuations, this allows one to fold the dynamics of the $Q$ space into effective rates which depend upon $\bf n_P$. This can be written as

\begin{eqnal}\label{eq:Qspace}
{\bf n_{Q}} = - {\bf M_{Q}}^{-1}(\bf M_{QP} n_P + \bf \Gamma_{Q})
\end{eqnal}
\begin{eqnal}\label{eq:Pspace}
\frac{d}{dt}{\bf n_{P}} &= (\bf M_{P} - M_{PQ}M_Q^{-1}M_{QP})n_P \\ & - {\bf M_{PQ}M_Q^{-1}\Gamma_{Q}} + {\bf \Gamma_{P}} \\
&= {{\bf M}_{\text{eff}}}{\bf n_P} + {{\bf \Gamma}_{\text{eff}}}
\end{eqnal}

But the applicability of such a separation in dynamical space needs to be quantitatively tested. In particular, for the constructed CR model to be applicable, it should satisfy Greenland's criteria \cite{Greenland_CR,Greenland_full,STOTLER_2007}, which requires evaluating the normalized eigenvalues and eigenvectors of ${\bf M}(n_e,T_e)$. The $N$ eigenvectors are arranged as the columns of an $N \times N$ matrix $\bf T$, in order of increasing eigenvalue, $\lambda$, and can be partitioned into 4 submatrices:

\begin{eqnal}\label{eq:Tspace}
{\bf T} =
\begin{bmatrix}
   {\bf T_{P}} & {\bf T_{PQ}} \\
   {\bf T_{QP}} & {\bf T_{Q}}
\end{bmatrix}
\end{eqnal}

In terms of these quantities, Greenland's criteria require that (i) $\lvert \lvert {\bf T_{QP}} \rvert \rvert \ll 1$ and (ii) $\lvert \lvert {\bf T_{QP}} {\bf T_{P}^{-1}} \rvert \rvert \ll 1$. From this point onwards, we will adopt in our evaluation an $N_P = 1$ CR model where the $P$ space consists of only the ground state for atomic helium being dynamically evolved. In this formulation, meta-stable species (e.g. $2^1S$, $2^3S$) are taken to be in steady state. Greenland's criteria for the $N_P = 1$ CR theory were previously examined in a range of conditions relevant to fusion plasmas and found to widely satisfy (i) and (ii) \cite{STOTLER_2007}, but there is an additional unresolved practical condition: (iii) the shortest timescales over which $P$ space states are evolved should be larger than the inverse of the smallest $Q$ space eigenvalue, i.e. $\frac{\partial}{\partial t} < \lvert \lambda_Q \rvert$. In more concrete terms, phenomena on timescales faster than $\tau_Q \equiv 1/\lvert \lambda_Q \rvert$ are not resolved. As a result, $\tau_Q$ represents the slowest timescale in $Q$ space, which is not tracked, and the ground state of atomic helium should be evolved on timescales slower than $\tau_Q$ for the separation of the two dynamical spaces to be consistent since all timescales faster than $\tau_Q$ are effectively instantaneous. For the CR formulation to be subsequently applicable in the spectroscopic analysis of plasma turbulence, the autocorrelation time of $n_e$ ($\tau_{n_e}$) and $T_e$ ($\tau_{T_e}$) must be larger than $\tau_Q$. Additionally, the exposure time of the experimental imaging diagnostic, $\tau_{GPI}$, should satisfy the timescale criterion of 

\begin{figure}[ht]
\includegraphics[width=1.0\linewidth]{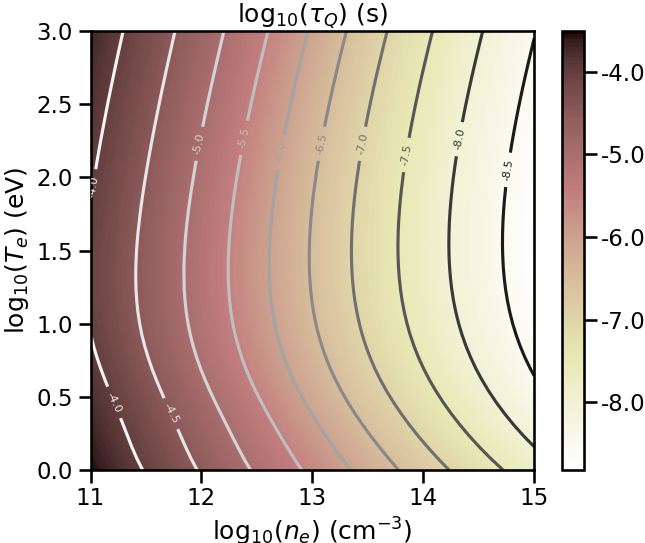}
\caption{\label{tauQ_5876}A contour plot of $\tau_Q$ for the $N_P = 1$ CR model scanned over a range of relevant electron densities and temperatures for magnetically-confined fusion plasmas. A logarithmic scale is applied on all axes including the colourbar.}
\end{figure}

\begin{eqnal}\label{eq:tauQ_validity}
\tau_Q < \tau_{GPI} < \tau_{n_e},\tau_{T_e}
\end{eqnal}

for consistency. This ensures the experimentally observed line emission in a single exposure time is based upon neutrals nominally excited by a unique $n_e$ and $T_e$ instead of a range of contributing magnitudes. Using revised cross-sections from \cite{RALCHENKO2008,Zholobenko_2018}, we report $\tau_Q$ under the $N_P = 1$ CR formulation in Figure \ref{tauQ_5876} at a range of $n_e$ and $T_e$ relevant to fusion plasmas. This quantity demarcates the temporal domain of validity. An important trend from the plot is that as $n_e$ increases, the limit on the temporal resolution of turbulent fluctuation measurements improves. For high plasma density fluctuations such as coherent filamentary structures, the resolution is roughly $\tau_Q \lesssim 1 \ \mu\text{s}$ for even $n_e \sim 10^{13} \ \text{cm}^{-3}$. As $n_e$ increases in higher field devices, the theoretical limit for resolving temporal scales improves. This aids the application of our GPI analysis for studying plasma turbulence in new regimes on upcoming tokamaks. A lower limit on spatial resolution for turbulence diagnostic imaging is set by $v_{HeI}/A(3^3D, 2^3P)$, provided that it is shorter than $v_{HeI} \tau_{n_e}$---or $v_{HeI} \tau_{T_e}$, if smaller---where $v_{HeI}$ is the particulate velocity of the atomic helium. The validity criteria for the $N_P = 1$ CR formulation are generally satisfied in analyzing the $3^3D \rightarrow 2^3P$ transition for fusion plasmas of sufficient density, but one should take care when checking validity in different scenarios. For example, if applying CR theory to cameras imaging different electronic transitions (e.g. for analysis of line ratios \cite{HeI_line_ratio1,HeI_line_ratio2,Griener_2018}) with long exposure times where $\tau_{n_e}, \tau_{T_e} < \tau_{GPI}$, the formulated CR theory is technically invalid as the condition given by Eq. \eqref{eq:tauQ_validity} is no longer met. This could potentially cause misalignment of $n_e$ and $T_e$ profiles when comparing existing experimental diagnostics towards closed flux surfaces, where plasma fluctuations are temporally faster than the observed autocorrelation time of far SOL turbulence \cite{labombard_evidence_2005}. Farther in the SOL as the plasma pressure drops, one should also be careful to check that $\tau_Q < \tau_{n_e}, \tau_{T_e}$. Diagnosing edge fluctuations thus necessitates sufficiently high resolution for both the experimental diagnostic and applied CR theory.

The $N_P = 1$ CR formulation permits any excited state population density in $Q$ space to be written as

\begin{eqnal}\label{eq:nQ_CR}
{\bf n_Q}\lvert_q = n(q) = R_0(q)n_en_i + R_1(q)n_en(1^1S)
\end{eqnal}

where $R_0(q)$ and $R_1(q)$ are known as population coefficients associated with recombination and electron impact physics. The temporal evolution of the ground state, the only species in $P$ space for this CR model, follows 

\begin{eqnal}\label{eq:nP_CR}
\frac{d}{dt}{\bf n_P} = \frac{d}{dt}{n(1^1S)} = \alpha_{CR}n_en_i - S_{CR}n_en(1^1S)
\end{eqnal}

where $\alpha_{CR}$ and $S_{CR}$ are the recombination and ionization rate coefficients, respectively. To generate photon emissivity coefficients from this CR model, Eq. \eqref{eq:nQ_CR} is multiplied by the Einstein A coefficient for the given radiative transition. For the 587.6 nm line, $A(3^3D, 2^3P) = 2 \times 10^7 \ {\text{s}}^{-1}$. If $q = 3^3 D$, by multiplying Eq. \eqref{eq:nQ_CR} with the corresponding spontaneous decay rate, one can compute

\begin{eqnal}\label{eq:PECexc}
\text{PEC}^{exc} = R_1(3^3D) A(3^3D, 2^3P)
\end{eqnal}
\begin{eqnal}\label{eq:PECrec}
\text{PEC}^{rec} = R_0(3^3D) A(3^3D, 2^3P)
\end{eqnal}

\begin{figure*}[ht]
\includegraphics[width=1.0\linewidth]{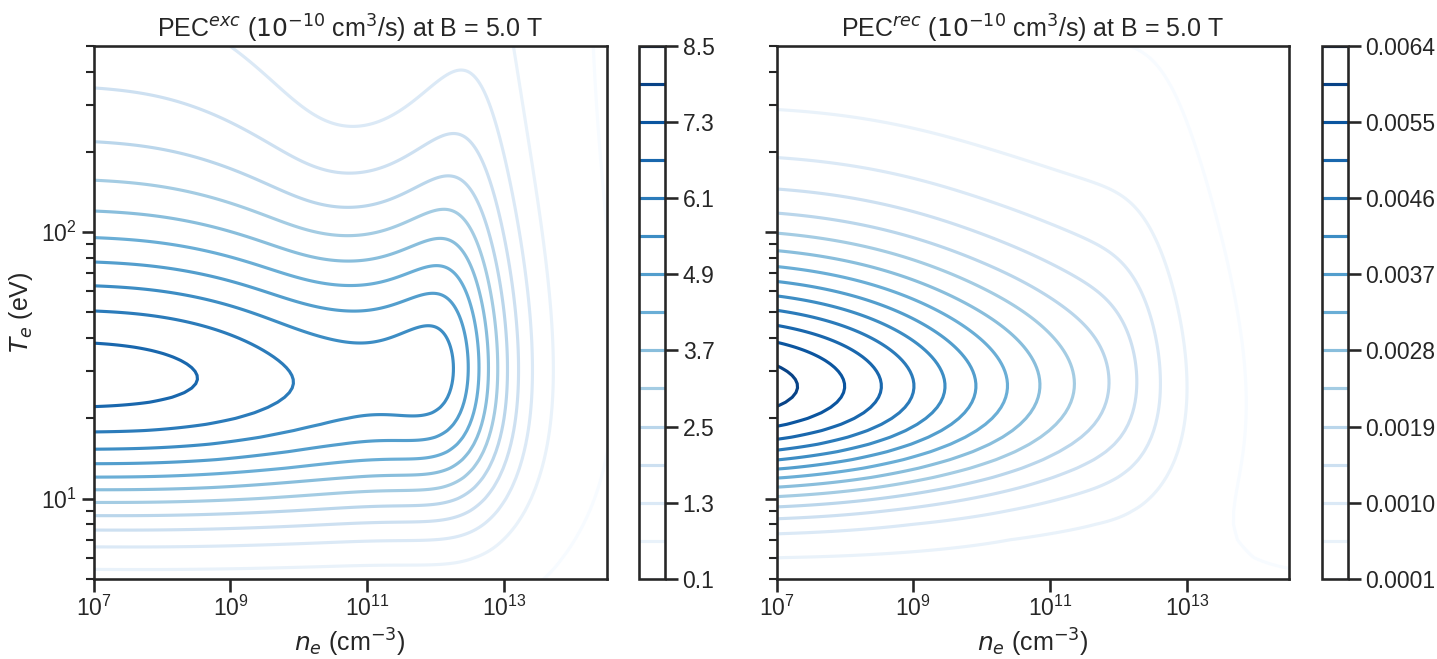}
\caption{\label{CR_rate3}Photon emissivity coefficients for the HeI 587.6 nm line based upon electron impact excitation (left) and recombination (right). These quantities are derived from the $N_P = 1$ CR model's population coefficients and follow Eqs. \eqref{eq:PECexc} and \eqref{eq:PECrec}.}
\end{figure*}
\begin{figure*}[ht]
\includegraphics[width=1.0\linewidth]{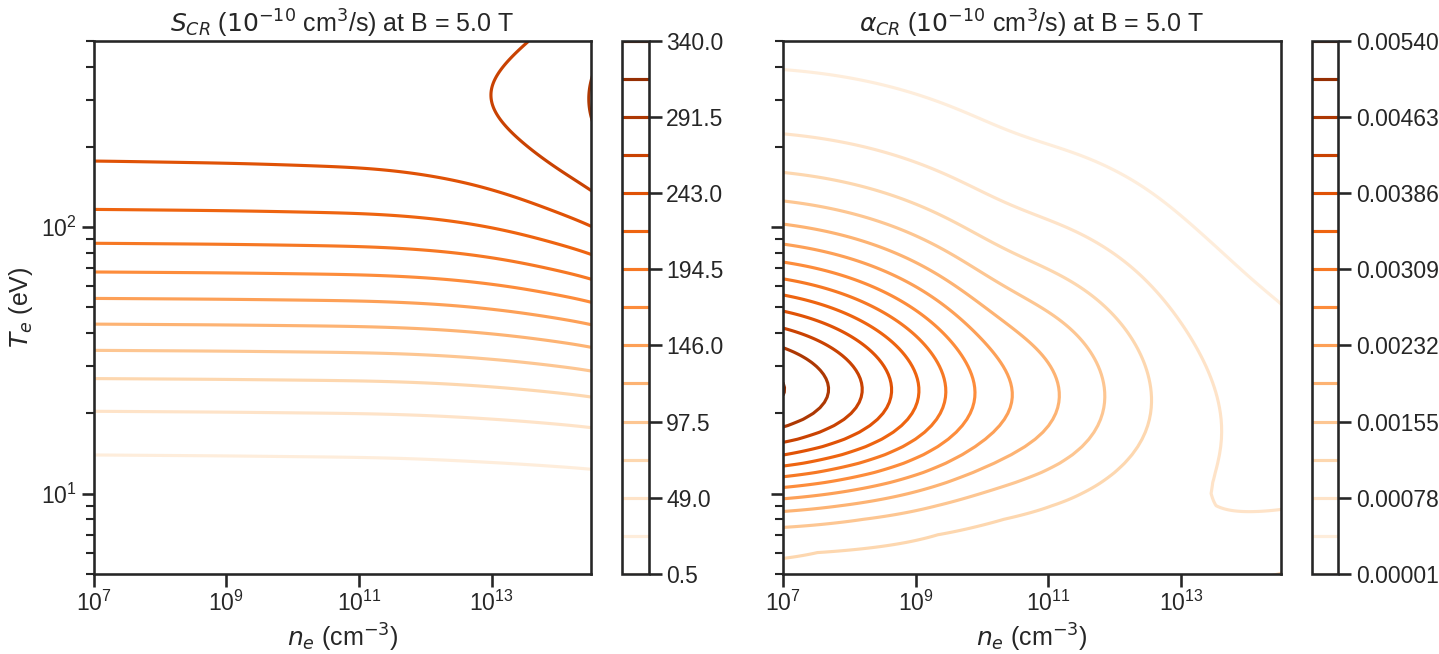}
\caption{\label{CR_rate5}Ionization (left) and recombination (right) rate coefficients derived from the $N_P = 1$ CR model. These quantities represent sinks and sources in Eq. \eqref{eq:nP_CR} for atomic helium when considering their transport in fusion plasmas.}
\end{figure*}

Contours of all coefficients along with their dependence on $n_e$ and $T_e$ are visualized at a magnetic field of $B = 5 \ \text{T}$ in Figures \ref{CR_rate3} and \ref{CR_rate5}. Given these rates, one can further simplify the expressions for Eqs. \eqref{eq:nQ_CR} and \eqref{eq:nP_CR} when modelling 587.6 nm line emission in the presence of edge plasma turbulence by removing the effects of volumetric recombination, i.e. $\text{PEC}^{exc} \gg \text{PEC}^{rec}$ and $S_{CR} \gg \alpha_{CR}$, which are negligible for edge fusion plasmas unless $n_i \gg n_0 \equiv n(1^1S)$, i.e. only if the HeII density is far greater than the ground state neutral helium density. The effects of charge-exchange are also neglected as the reaction rate is small compared to electron impact ionization for atomic helium as long as $5 \ \text{eV} < T_e < 5 \ \text{keV}$ \cite{AMJUEL}. Note that this is not necessarily true for other atomic or molecular species, e.g. deuterium \cite{Helander_Catto_K1994}, but allows for an expression of 587.6 nm photon emissivity given by

\begin{eqnal}\label{eq:emissivity_GPI}
I = C n_0 n_e \text{PEC}^{exc}(n_e,T_e) = Cn_0 f(n_e,T_e)
\end{eqnal}

where $f(n_e,T_e)$ can be interpreted as the photon emission rate per neutral consistent with the $N_P = 1$ CR model. Using an oft-applied exponential model of $f(n_e,T_e) \propto n_e^{\alpha_n} T_e^{\alpha_T}$ and treating $\alpha_n$ and $\alpha_T$ as constants could yield erroneous emissivity predictions where fluctuations of order unity are beyond the perturbative regime. It is important to therefore retain the full range of dependency on $n_e$ and $T_e$. A constant factor $C$ is introduced in \eqref{eq:emissivity_GPI} to account for calibration of the instrument used to measure the line radiation, including effects introduced by the finite thickness of the  observed emission cloud.

\section{\label{sec:level3}Experimental turbulence imaging of helium line emission}
Our experimental analysis technique is generic to any plasma discharge on Alcator C-Mod where good fast camera data exist for the 587.6 nm line. The plasma discharge chosen for this work constitutes 1120711021. This is a majority deuterium, lower single null diverted ohmic plasma with an on-axis toroidal magnetic field of 5.4 T and plasma current of 0.83 MA. The tokamak itself has a major radius of 0.68 m and minor radius of 0.22 m. The discharge has independent diagnostic measurements from a main chamber scanning probe equipped with a mirror Langmuir probe (MLP) biasing system run in a swept mode in the edge plasma \cite{MLP_origin}. Based on Thomson scattering and electron cyclotron emission diagnostic measurements, the core electron density and temperature are $2.0 \times 10^{20} \ \text{m}^{-3}$ and $1.5$ keV, respectively.


For the present work, the GPI diagnostic on the Alcator C-Mod tokamak \cite{2002_Zweben,GPImanual} was configured to capture visible light at a wavelength of 587.6 nm arising from the dynamic interaction of edge plasma turbulence with neutral helium puffed locally to the imaged region. This is a commonly used technique akin to other plasma diagnostics such as beam emission spectroscopy (BES) \cite{BES_McKee}. Helium is an ideal choice for 2-dimensional turbulence imaging for several reasons: its low atomic number results in radiative losses minimally perturbing the plasma state; its larger ionization energy allows for greater neutral penetration than thermal deuterium; its lack of molecular interactions reduces complexity in modelling; and its neutrality keeps its transport independent of external magnetic fields. The localized distribution of atomic helium also provides a greater contrast to the background emissivity in fusion plasmas which primarily fuel with hydrogen isotopes. HeI emission was imaged onto a Phantom 710 fast camera, installed on Alcator C-Mod in 2009 to view the outboard midplane region \cite{GPImanual}. The camera has a maximum framing rate of 400,000 frames/s at 2.1 $\mu$s-exposure/frame when 64 $\times$ 64 pixels are being read out, and each pixel is approximately $20 \ \mu$m $\times \ 20 \ \mu$m. The diagnostic's resultant temporal resolution is 2.5 $\mu$s as it takes 0.4 $\mu$s to read values from the pixel array. The fast camera has a built-in positive offset of approximately 80 counts, which is subtracted from all GPI signals before analysis of the experimental data \cite{GPImanual}. Based upon the manufacturer’s specifications and sample bench tests, the fast camera measurements are expected to vary linearly with light level over the pixels analyzed. The brightness is thus offset in absolute magnitude by a constant scale factor and accounted for in the framework.

\begin{figure}[ht]
\includegraphics[width=1.0\linewidth]{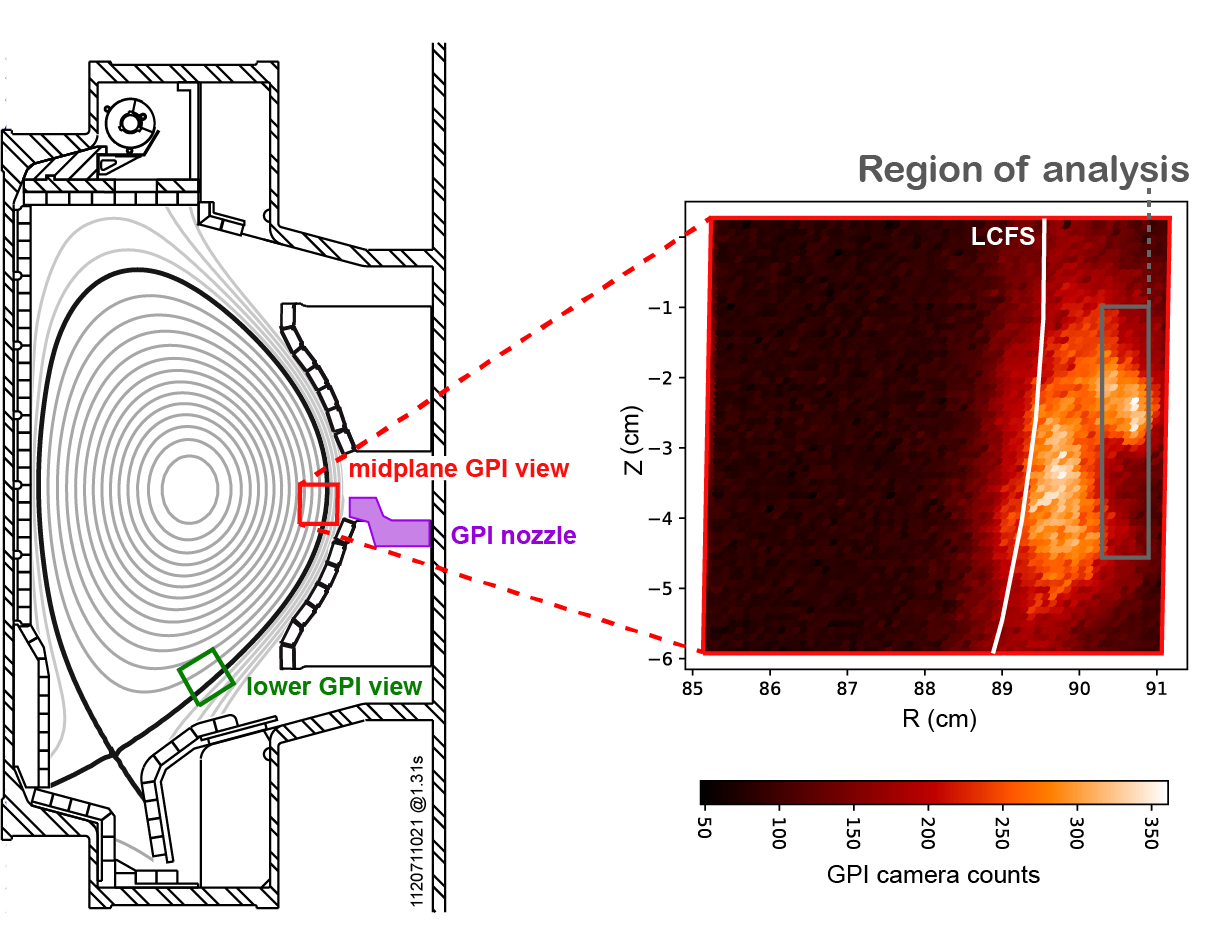}
\caption{\label{CMOD_GPI}Visualization of the experimental GPI setup on a poloidal cross section of a lower single null diverted plasma discharge (1120711021) on Alcator C-Mod. In this manuscript, we use measurements from the midplane fast camera with a 587.6 nm optical filter with full width at half maximum of 11.4 nm which has a largely field-aligned view of edge fluctuations. The expansion at right shows raw counts measured by the fast camera at $t = 1.312858$ s, and includes overlays of both the last closed flux surface and the approximate domain of the analysis described in Section \ref{sec:level5}.}
\end{figure}

\begin{figure}[ht]
\includegraphics[width=1.0\linewidth]{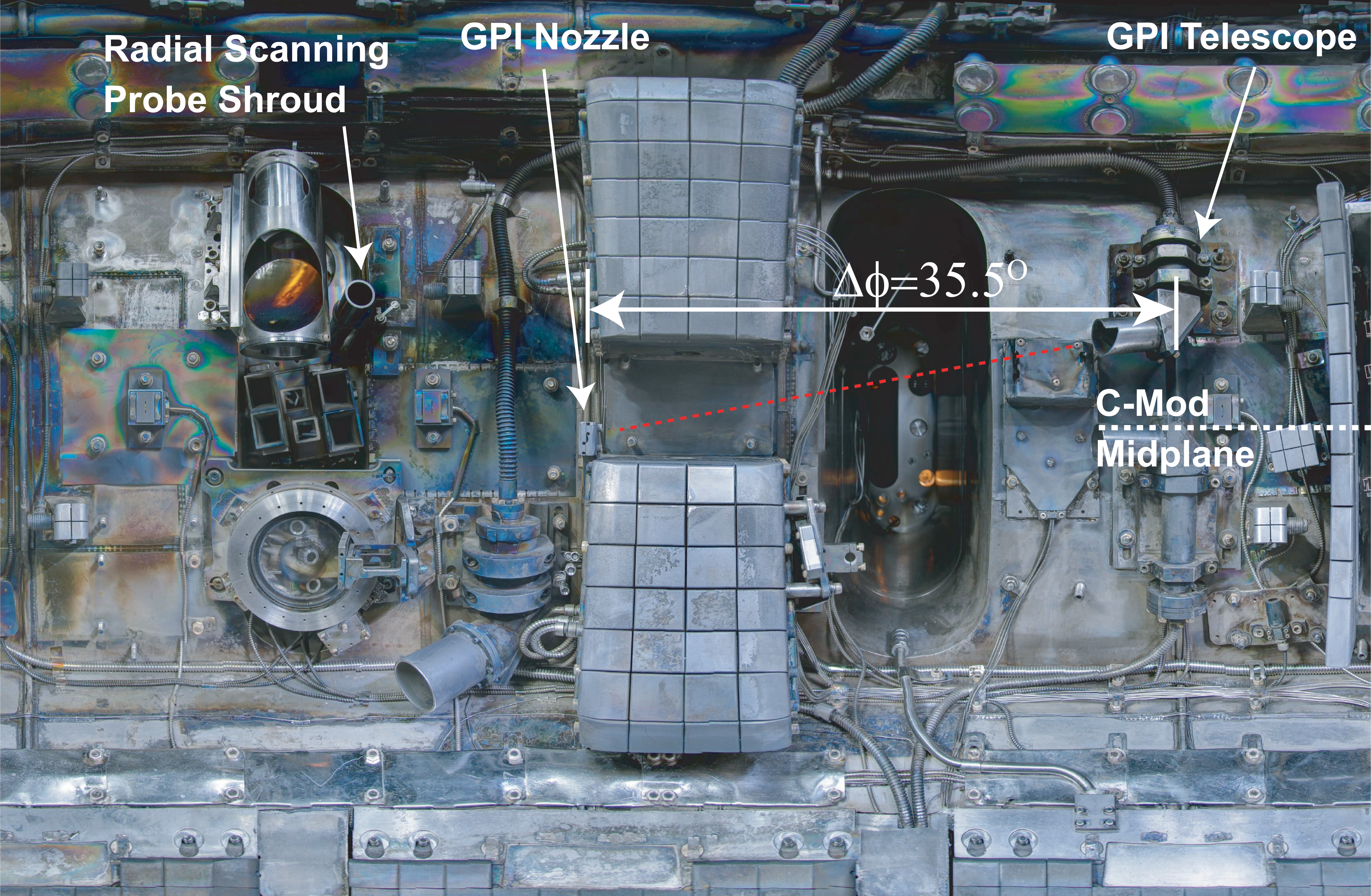}
\caption{\label{CMOD_Phantom}A section of a panoramic photo of the Alcator C-Mod outer wall, showing approximately one quarter of the device and centered on the split poloidal limiter next to the gas puff imaging measurement location. Labels indicate the position of the GPI nozzle, its imaging telescope, and an approximate line of sight (red dashed). The position of the radial scanning probe, which provides the mirror Langmuir probe measurements, is also exhibited.}
\end{figure}

A coherent fiber bundle/image guide was used to couple light from viewing optics mounted on the outer wall of the vacuum vessel to the Phantom camera detector array. The optics imaged a roughly 60 mm $\times$ 60 mm region in the $(R,Z)$-plane just in front of a gas puff nozzle through a vacuum window onto the image guide. The viewing chords pointed downwards at a fixed angle of $11.0^{\circ}$ below horizontal towards the vertically-stacked 4-hole gas nozzle displaced from the telescope by approximately $35.5^{\circ}$ in toroidal angle. The central ray of the imaged view thus pierced the gas puff plane approximately parallel with the local magnetic field line \cite{GPImanual}. This aligns the GPI optics with field-aligned fluctuations for typical operational parameters of an on-axis toroidal field of 5.4 T and plasma current of 1.0 MA. For discharge 1120711021 conducted at a plasma current of 0.83 MA, the viewing chords are oriented at an angle of approximately 2$^{\circ}$ to the local field. Spatial blurring due to this angular misalignment, $\theta_B$, consequently limit resolution to $\Delta x = L_{||} \tan \theta_B$, where $L_{||}$ is the emission cloud's length parallel to the local magnetic field line. For $L_{||}$ between 5 -- 40 mm, the smearing will be 0.2 -- 1.4 mm in addition to the 1 mm pixel spot size in the image plane. Since the gas cloud expands after exiting the 4-hole nozzle and the local magnetic field's pitch angle varies, the smearing increases for those chords farther away from the nozzle depending upon the collimation of the gas cloud \cite{Zweben_2009,Zweben_2017}. With this setup and under these plasma conditions, we thus estimate the spatial resolution over the portion of the field-of-view that we analyze to be approximately 1-2 mm. A visualization of the experimental setup's poloidal cross-section is displayed in Figure \ref{CMOD_GPI} with the camera telescope in Figure \ref{CMOD_Phantom}.


As noted above, helium gas is injected into the vessel via four vertically-displaced plasma-facing capillaries located at $Z = -4.2, -3.4, -2.6, \text{and} -1.9$ cm, which are mounted in a port on a shelf just below the outer midplane sitting in the shadow of two outboard limiters. The position $Z = 0$ corresponds to the vertical location of the magnetic axis. The gas tubes' orifices are positioned at $R = 91.94$ cm with the channel exit diameter being 3 mm. The helium atoms are supplied by the Neutral gas INJection Array (NINJA) storage and delivery system \cite{NINJA_thesis} which has a pneumatically-controlled valve at the plenum which is connected to a 3.48-m-long, 1-mm-diameter capillary that feeds the 4 diverging gas tubes. Previous measurements indicate that the gas cloud exiting a single 1-mm-diameter capillary expands with angular half-width of 25$^{\circ}$ in both the poloidal and toroidal directions. This is the basis for our estimate of the 1--2 mm spatial resolution given above. Due to the tubes' spatial displacement, the helium gas puff is intended to be relatively uniform in the vertical direction. By definition, there is a shock at (or near) the vacuum-nozzle interface for this sonic flow since only particles moving downstream can escape and there is consequently no information being communicated to upstream particles \cite{shocks_in_tubes_Parks_and_Wu}. The neutral dynamics thus transition from a fluid regime in the gas tube to a kinetic regime upon entering the tokamak from the nozzle. The HeI exiting the diverging nozzles is approximately modelled by a drifting, cut-off Maxwellian distribution with a mean radial velocity of $v_x \sim -900$ m/s and a mean vertical velocity of $v_y \sim -20$ m/s since the direction of the non-choked flow in the gas tubes is roughly 2.4$^{\circ}$ away from being purely radial in orientation. 


\begin{figure*}[ht]
\includegraphics[width=1.0\linewidth]{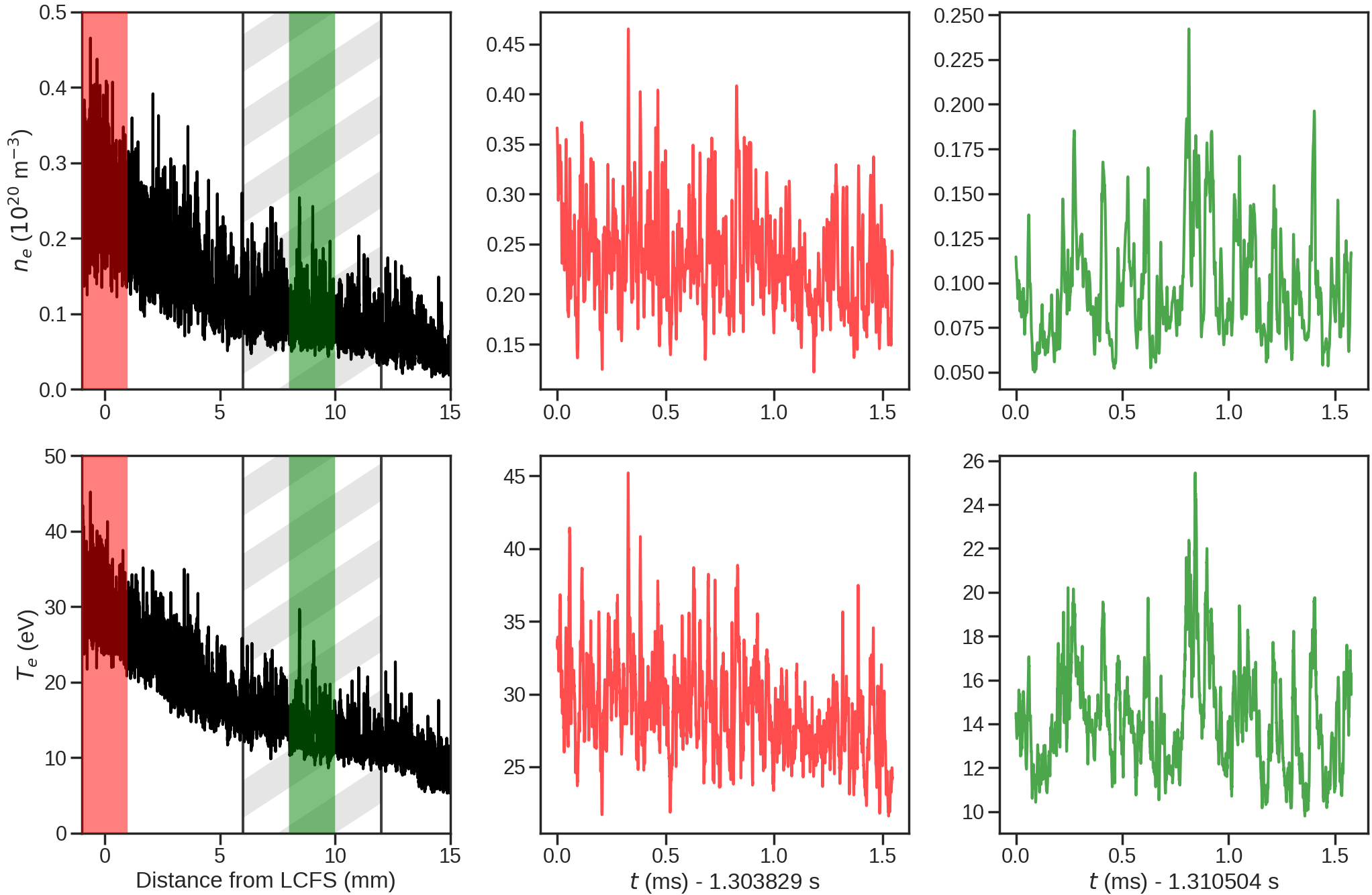}
\caption{\label{probe_1120711021} Experimental $n_e$ and $T_e$ measurements in discharge 1120711021 from the independent scanning mirror Langmuir probe. The full probe plunge duration is $1.288 < t \ \text{(s)} < 1.318$. As the probe is scanning back from closed flux surfaces, time series of $n_e$ and $T_e$ are plotted to compute autocorrelation times. At distances near the LCFS (red), $\tau_{n_e}$ and $\tau_{T_e}$ are approximately 4.2 $\mu$s and 6.1 $\mu$s, respectively. Farther into the SOL (green), $\tau_{n_e}$ and $\tau_{T_e}$ increase to 15.6 $\mu$s and 22.9 $\mu$s, respectively. For reference, the gray shaded region roughly corresponds to the radial extent of the GPI data analyzed.}
\end{figure*}

To examine the experimental relevance of applying the $N_P = 1$ CR theory outlined in Section \ref{sec:level2} for analysis of edge plasma turbulence on Alcator C-Mod, we review a few key characteristic parameters of interest based upon scanning MLP measurements of $n_e$ and $T_e$ in plasma discharge 1120711021. Magnetically disconnected from the GPI field of view, the scanning MLP is located at $Z = 11.1$ cm roughly $20^{\circ}$ in toroidal angle from the GPI view and radially traverses the tokamak plasma from the far edge to just inside the last closed flux surface (LCFS) with a temporal resolution of 0.3 $\mu$s. Measurements mapped to the midplane radius are visualized in Figure \ref{probe_1120711021} based upon a probe plunge nearly coincident temporally with our GPI analysis of this plasma discharge. While the probe bias is inherently perturbative due to the collection of charged particles, we assume that its effects on local plasma conditions are negligible \cite{kuang_thesis}. From the MLP data, we can obtain autocorrelation times of fluctuations near the LCFS and approximately 8 - 10 mm radially outward into the SOL when mapped to the midplane radius. Towards closed flux surfaces, $\tau_{n_e}$ and $\tau_{T_e}$ are approximately 4.2 $\mu$s and 6.1 $\mu$s, respectively. In the far SOL, $\tau_{n_e}$ and $\tau_{T_e}$ increase to 15.6 $\mu$s and 22.9 $\mu$s, respectively. Since the probe has a finite velocity and the autocorrelation length of fluctuations is finite, these estimates of $\tau_{n_e}$ and $\tau_{T_e}$ act as conservative lower bounds as long as there is no aliasing nor phase-alignment between the probe's motion and turbulence structures. We note that the fast camera exposure time of 2.1 $\mu$s is expected to be suitable for analysis of edge plasma fluctuations in this ohmic discharge, although faster cameras could be helpful in analyzing plasma conditions. Further, for turbulence near the LCFS where $n_e \gtrsim 10^{19} \ \text{m}^{-3}$  and $T_e \gtrsim 20$ eV, then $\tau_Q < 1 \ \mu$s, and the condition of $\tau_Q < \tau_{exp} < \tau_{n_e},\tau_{T_e}$ is well-satisfied. For fluctuations farther out into the SOL, this condition is still generally valid especially in the treatment of high pressure filaments, but one should be careful when $n_e$ drops below $2.5 \times 10^{18} \ \text{m}^{-3}$ in fusion plasmas. For spectroscopic techniques analyzing line intensities, each optical camera's exposure time needs to be suitably adjusted to satisfy the timescale condition. This is especially important towards closed flux surfaces where long camera exposure periods and shorter autocorrelation times would render time-dependent examination of brightness ratios arising from turbulent fluctuations as inconsistent. 

Our framework outlined in the next sections can be applied to regions with arbitrary geometries (e.g. X-point, divertor) if using sufficiently planar helium beams where the width of the collimated gas is smaller than the parallel autocorrelation length of the plasma fluctuations in the direction of the viewing chords. Since the viewing chords are roughly field-aligned over the pixels being analyzed, this parallel scale condition is expected to be satisfied. Finally, we note that the signal-to-noise ratio degrades in the inboard portion of the field-of-view, which includes plasma close to or on closed flux surfaces where the electron pressure and ionization rate increase sharply \cite{jwhughes1,jwhughes2}. Accordingly, we analyze fluctuations a few millimetres away from the LCFS on a 2-dimensional $(R,Z)$-grid co-located at the nominal gas puff plane. In future work, if greater neutral penetration can be achieved such that high signal-to-noise can be attained on closed flux surfaces, the capability to then probe pedestal dynamics also exists where priming our optimization framework on available 1-dimensional data may help \cite{Mathews2020}. This opportunity may already be viable on devices with smaller  line-integrated $n_e$ and the methodology can extend to unmagnetized plasmas, too. 




\section{\label{sec:level4}Deep learning of time-dependent neutral transport physics and collisional radiative theory}

Combining the theory governing atomic emission and neutral transport with experimental turbulence measurements via fast camera imaging into an integrated analysis framework requires sufficiently sophisticated modelling techniques. Neural networks are differentiable computational programs that can provide natural representations for physical systems determined by differential-algebraic equations. They extend ordinary regression models into robust universal function approximators with generalized constraints that are highly effective in solving inverse optimization problems with sufficient training given their high plasticity. We outline a novel multi-network deep learning framework custom-built for analysis of 587.6 nm helium line emission in fusion plasmas to uncover $n_e$, $T_e$, and $n_0$. The networks only receive experimental brightness measurements from GPI while being optimized against the $N_P = 1$ CR theory for photon emissivity along with the continuity equation for neutral transport which accounts for ionization of helium atoms on turbulent scales. In this way, we combine training upon both mathematical laws and observational data. To begin, we represent the unobserved quantities $n_e$, $T_e$, and $n_0$ each with their own neural network. The initial layer inputs correspond to the local spatiotemporal points $(x,y,t)$, with the $(x,y)$-coordinate being equivalent to $(R,Z)$, from the approximately 2-dimensional domain viewed by the fast camera in the poloidal plane of the gas puff nozzle. The only output of each network is the respective dynamical variable being represented. Every network's inner architecture consists of 5 hidden layers with 150 neurons per hidden layer and hyperbolic tangent activation functions ($\sigma$) using Xavier initialization \cite{GlorotAISTATS2010}. To provide reasonable intervals for the optimization bounds, the networks for $n_e$, $T_e$, and $n_0$ are constrained via output activation functions to be between $2.5 \times 10^{18} < n_e \ (\text{m}^{-3}) < 7.5 \times 10^{19}$, $2.5 < T_e \ (\text{eV}) < 150.0$, and $0.1 < n_0 \ (\text{arb. units}) < 10.0$. While required for numerical stability, care must be taken since solutions existing near these limits may not be fully converged. We note that the learnt constant calibration factor, $C$, is similarly represented by a network but does not vary spatially nor temporally. Physically, this results in $n_0$ being determined up to a constant scaling. The scalar constant also accounts for the 2-dimensional approximation of the localized gas puff, which has a finite toroidal width from the helium atoms exiting the capillaries. By assuming $n_e$, $T_e$, and $n_0$ to be roughly uniform along the camera's sightline, the effect of this finite volume is absorbed when learning the calibration factor. While the 2-dimensional approximation is reasonable for sufficiently planar gas injection, the deep framework can technically be generalized towards natively handling 3-dimensional space since it employs a continuous domain without any discretization. This is a future extension.

\begin{figure}[ht]
\includegraphics[width=1.0\linewidth]{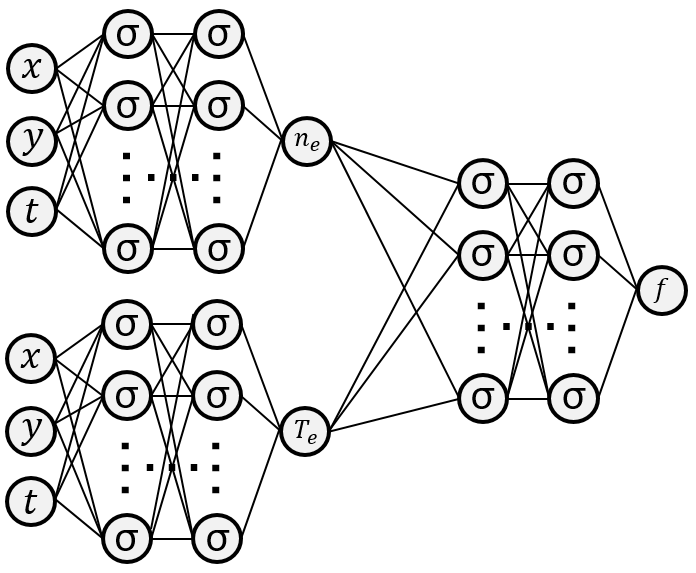}
\caption{\label{network_structure_f} Structure of networks to represent $f(n_e, T_e) = n_e \text{PEC}^{exc}(n_e, T_e)$ which is one of the terms composing the total emissivity function, $I = C n_0 f(n_e, T_e)$. The ionization rate per neutral, $n_e S_{CR}(n_e, T_e)$, is similarly represented when applied in the transport equation and important to account for ``shadowing" of neutrals \cite{2002_Zweben,STOTLER2003,Wersal_2017}. The left side of the overall network consists of the networks for the predicted $n_e$ and $T_e$, while the right side output represents the photon emissivity per neutral given by the $N_P = 1$ CR theory.}
\end{figure}

Our optimization is conducted in stages. To begin learning CR theory, we construct novel neural network structures where the outputs of the $n_e$ and $T_e$ networks serve as inputs to a new architecture representing the photon emissivity per neutral, $f \equiv f(n_e, T_e)$. The connectivity of the neurons conjoining $n_e$ and $T_e$ towards the network's output, $f$, is visualized in Figure \ref{network_structure_f}. These weights and biases are trained against $n_e \text{PEC}^{exc}(n_e, T_e)$, which is derived from the $N_P = 1$ CR theory. The corresponding emissivity coefficient is plotted in Figure \ref{CR_rate3}. The ionization rate per neutral, $n_e S_{CR}(n_e, T_e)$, which is based upon the coefficient plotted in Figure \ref{CR_rate5}, is similarly represented by an architecture with $n_e$ and $T_e$ serving as inputs. All this training of the two architectures representing $f(n_e, T_e)$ and $n_e S_{CR}(n_e, T_e)$ is conducted in the first stage prior to any optimization against the fast camera data. This ensures the next stages involving training with collisional radiative constraints take place under an integrated optimization framework with all quantities being represented by neural networks. For numerical purposes, $n_e$ are $T_e$ are normalized by $10^{19} \ \text{m}^{-3}$ and $50 \ \text{eV}$, respectively, and time is converted to units of microseconds during the optimization. For low temperature plasmas where $T_e < 2$ eV, training with the networks and output CR coefficients from \cite{GOTO_2003,Zholobenko_2021} based upon fitted electron impact cross-sections should be carefully checked due to potential corrections to fits for collision strengths at such low energies. We also prime only the $n_e$ and $T_e$ networks against constants of $10^{19} \ \text{m}^{-3}$ and 50 eV, respectively, for initialization during this first stage. The priming of $n_e$ and $T_e$ and learning of CR coefficients by their respective networks takes place over the first 5 of 20 total hours of training on 32 cores with Intel Haswell-EP processors. 

Next, we train the $n_e$, $T_e$, and $C$ networks against Eq. \eqref{eq:emissivity_GPI} such that that the predicted brightness intensity consistent with CR theory matches experimental measurements from the fast camera. Based upon past high-resolution MLP data, turbulent fluctuations propagating in the edge were observed to exhibit strong correlations between the electron density and electron temperature \cite{KUBE2019_neTecorr}. Additional constraints are thus placed in our optimizer such that solutions where $n_e$ and $T_e$ are correlated are favoured. This helps to avoid the learning of trivial solutions. Namely, the full loss function being collectively trained upon in this second stage is

\begin{eqnal}\label{eq:loss_GPI}
\mathcal{L}^{C,n_e,T_e} &= \frac{1}{N_0}\sum_{i=1}^{N_0} (\mathcal{L}_{GPI} + C_1 \mathcal{L}_{corr} + C_2 \mathcal{L}_{relcorr}),
\end{eqnal}

where

\begin{eqnal}\label{eq:loss_GPI1}
\mathcal{L}_{GPI} &= \lvert I^*(x^i_0,y^i_0,t^i_0) - I_0 \rvert^2 
\end{eqnal}

\begin{eqnal}\label{eq:loss_corr}
\mathcal{L}_{corr} &= - [n^*_e(x^i_0,y^i_0,t^i_0) - \langle n^*_e(x^i_0,y^i_0,t^i_0) \rangle] \times \\& [T^*_e(x^i_0,y^i_0,t^i_0) - \langle T^*_e(x^i_0,y^i_0,t^i_0) \rangle]
\end{eqnal}

\begin{eqnal}\label{eq:loss_relcorr}
\mathcal{L}_{relcorr} &= \frac{\lvert \langle n^*_e(x^i_0,y^i_0,t^i_0) - \langle n^*_e(x^i_0,y^i_0,t^i_0) \rangle \rangle \rvert^2}{\lvert \langle T^*_e(x^i_0,y^i_0,t^i_0) - \langle T^*_e(x^i_0,y^i_0,t^i_0) \rangle \rangle \rvert^2} \\&+ \frac{\lvert \langle T^*_e(x^i_0,y^i_0,t^i_0) - \langle T^*_e(x^i_0,y^i_0,t^i_0) \rangle \rangle \rvert^2}{\lvert \langle n^*_e(x^i_0,y^i_0,t^i_0) - \langle n^*_e(x^i_0,y^i_0,t^i_0) \rangle \rangle \rvert^2}
\end{eqnal}

with $I^*(x^i_0,y^i_0,t^i_0)$ following Eq. \eqref{eq:emissivity_GPI}, and the points $\lbrace x_0^i,y_0^i,t_0^i,I_{0}^i\rbrace^{N_0}_{i=1}$ corresponding to the set of observed data from GPI. Here we use the notation that superscripts on $\mathcal{L}$ identify the multiple networks being simultaneously trained during optimization of the respective loss function, e.g. $\mathcal{L}^{C,n_e,T_e}$ indicates that the networks for $C$, $n_e$, and $T_e$ are being jointly optimized against this particular loss function. We note that the results from the converged solutions reported in Section \ref{sec:level5} are largely unchanged by removing \eqref{eq:loss_relcorr} in the optimization framework, although keeping it was found to enhance stability and thus the total number of realizations that converge. Better physics-informed optimization constraints may exist and should be investigated going forward to advance this turbulence analysis. While the coefficients $C_1$ and $C_2$ in Eq. \eqref{eq:loss_GPI} can be adaptively adjusted for optimal training in each iteration, they are set to constants of 1000 and 1, respectively, in this framework. The variables with asterisks symbolize predictions by their respective networks at the spatiotemporal points being evaluated during training. The notation $\langle X \rangle$ denotes the batch sample mean of $X$. This second training stage lasts for 100 minutes.

The next stage involves optimizing the $n_0$ network against both Eq. \eqref{eq:loss_GPI1} and its transport equation which accounts for neutral drifts and fluctuation-induced ionization of helium. Namely, in implicit form,

\begin{figure*}[ht]
\includegraphics[width=1.0\linewidth]{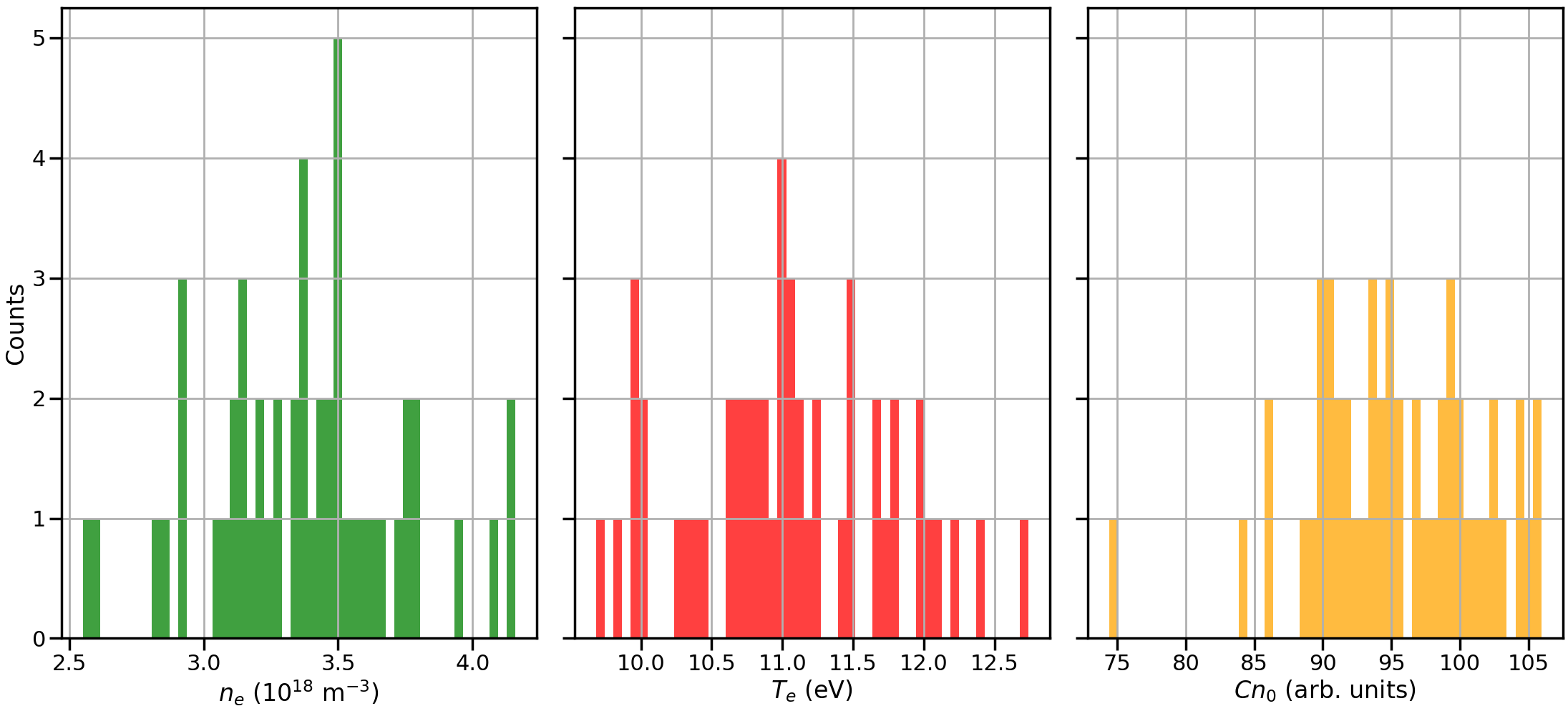}
\caption{\label{learned_1D_turbulence_histogram}Histogram of the $n_e$, $T_e$, and $Cn_0$ fluctuations at $[R = 90.7 \text{ cm}, Z = -4.2 \text{ cm}, t = 1.312850 \text{ s}]$, i.e. with corresponding normalized poloidal magnetic flux coordinate $\psi_n=1.07$, based upon 50 converged realizations when training against experimental GPI data from plasma discharge 1120711021 in the optimization framework. The ensemble mean and standard deviation of these realizations are used to construct the results presented in Section \ref{sec:level5}.}
\end{figure*}

\begin{eqnal}\label{eq:loss_fn0}
f_{n_0} = \frac{\partial n_0}{\partial t} + \frac{\partial (n_0 v_x)}{\partial x} +  \frac{\partial (n_0 v_y)}{\partial y} + n_0 n_e S_{CR}
\end{eqnal}

where, based upon 3-dimensional Monte Carlo neutral transport simulations of this region, closures of $v_x \sim -900$ m/s and $v_y \sim -20$ m/s are applied for modelling HeI as it exits the capillaries into the GPI frame of view \cite{SGBAEK_DEGAS2_GPI_CMOD}. This approximation of a drifting, cut-off Maxwellian for HeI may be reasonable for a narrow radial region, but the true velocity distribution characterizing helium gas particles becomes increasingly skewed the farther one goes away from the gas nozzles. Modelling other atomic and molecular species (e.g. deuterium) in this way may be inadequate as charge-exchange and recombination effects on trajectories are increasingly important. Also, neutral-neutral collisions and their impacts on velocity closures are presently neglected in this treatment. This allows for a scaling constant to be factored out of Eq. \eqref{eq:loss_fn0}, i.e. permitted by its linearity in $n_0$. If using a sufficiently high spectral resolution spectrometer to view the emission cloud, the Doppler shift can be experimentally measured. We leave this task for further exploring momentum transport physics and potentially even learning the velocity closure directly from the GPI data within our optimization framework for future work. 

The null formulation following Eq. \eqref{eq:loss_fn0} is vital for training since all physical terms collectively sum to zero when the unknown dynamical variables in the equation are correctly solved to self-consistently account for neutral propagation and ionization. The physical theory is computationally expressed by differentiating the $n_0$ neural network with respect to its input spatiotemporal coordinates via application of chain rule through automatic differentiation \cite{tensorflow2015-whitepaper}. By then multiplying and adding the graph outputs to construct representations of the physical constraints, the network for $n_0$ can be trained against \eqref{eq:loss_GPI} and \eqref{eq:loss_fn0} to satisfy the physical theory constraining the nonlinear connection between networks. This accounting of Eq. \eqref{eq:loss_fn0} is particularly essential since the Kubo number, which quantifies the strength of turbulent perturbations on neutral transport, is large ($\gtrsim 1$) for helium \cite{Kubo_original,Kubo_number,SGBAEK_DEGAS2_GPI_CMOD}. There are no explicit boundary conditions applied for $n_0$, but instead we train the network against the fast camera's experimentally measured intensities to learn how $n_0$ should be treated around the boundaries of the analyzed camera image. Namely, the loss function in this third following stage is given by

\begin{eqnal}\label{eq:loss_n0_train}
\mathcal{L}^{n_0} &= \frac{1}{N_0}\sum_{i=1}^{N_0} \mathcal{L}_{GPI} + \frac{C_{f_{n_0}}}{N_f}\sum_{j=1}^{N_f} \mathcal{L}_{f_{n_0}}
\end{eqnal}

with

\begin{eqnal}\label{eq:loss_fn0_train}
\mathcal{L}_{f_{n_0}} &=  \lvert f^*_{n_0}(x^j_f,y^j_f,t^j_f) \rvert^2 ,
\end{eqnal}

where $\lbrace x_f^j,y_f^j,t_f^j\rbrace^{N_f}_{j=1}$ denote the set of collocation points which can span any arbitrary domain but taken to be equivalent to the ones encompassed by $\lbrace x_0^i,y_0^i,t_0^i,I_{0}^i\rbrace^{N_0}_{i=1}$, and $f^*_{n_0}$ is the null partial differential equation prescribed by Eq. \eqref{eq:loss_fn0} in normalized form directly evaluated by the neural networks. For the remainder of the training time, Eqs. \eqref{eq:loss_GPI} and \eqref{eq:loss_n0_train} are sequentially trained in repeating intervals of 100 minutes to iteratively find convergence in their respective networks. The only difference in later stages is that $C$ is no longer a free parameter whilst training against Eq. \eqref{eq:loss_GPI}, and $C_{f_{n_0}}$ in Eq. \eqref{eq:loss_n0_train} is increased from $10^2$ to $10^6$ to improve the focused learning of neutral transport physics. 
If $C_{f_{n_0}}$ is increased any higher, we risk finding trivial solutions at a higher occurrence. Generalizing our training framework to adaptively update training coefficients \cite{wang2020understanding} is an important pathway for future investigation. All loss functions are optimized with mini-batch sampling where $N_0 = N_f = 1000$ using the L-BFGS algorithm---a quasi-Newton optimization algorithm \cite{10.5555/3112655.3112866}. Also, points found to have difficulty converging (e.g. optimizer becomes stuck in local minima) were removed from training in subsequent stages to improve learning of turbulent fluctuations in remaining regions of the spatiotemporal domain analyzed. In the end, this results in the multi-network framework training only on 8 (radial) $\times$ 38 (vertical) pixels over 39 frames imaged by the fast camera. We note that by embedding $f(n_e,T_e)$ in Figure \ref{network_structure_f}, the emissivity predictions by the networks are forced to satisfy CR theory. Similarly, the ionization rate per neutral, $n_e S_{CR}$, is encoded in Eq. \eqref{eq:loss_fn0}. We thus ensure that the unobserved $n_e$, $T_e$, and $n_0$ being learnt are in agreement with the experimentally measured brightness while trying to satisfy the neutral transport physics for HeI which self-consistently includes time-dependent ionization in the presence of plasma turbulence. The repeated differentiation and summation of networks to represent every term in the ascribed loss functions resultantly constructs a far deeper computation graph representing the collective constraints beyond the 8 hidden layers in each dynamical variable's individual network. The cumulative graph is therefore a truly deep approximation of the physics governing the observed 587.6 nm line emission.

Due to the stochastic nature of the initialization and multi-task training, learned solutions for $n_e$, $T_e$, $n_0$, and $C$ vary each time an individual optimization is run. This may arise due to a unique solution not necessarily existing given the above optimization constraints. Therefore, we run an ensemble of realizations and consider this collection of runs which roughly follow Gaussian statistics. Based upon past testing within our optimization framework, the necessary criteria for convergence in normalized units are set to $\mathcal{L}_{GPI} < 10^{2.5}$, $\mathcal{L}_{corr} < -10^3$, and $\mathcal{L}_{f_{n_0}} < 10^{-3}$. Checks for spurious gradients, trivial solutions, and a low number of training iterations were additionally investigated for downselecting converged realizations. For analysis of C-Mod discharge 1120711021, there were 800 runs with 50 sufficiently converging within our present analysis. The scatter in learned turbulent fluctuations among these realizations is used to quantify uncertainty intervals associated with the optimization framework, and as an example, the distribution of inferred measurements at a particular spatial and temporal point are plotted in Figure \ref{learned_1D_turbulence_histogram}. It is also important to note that the loss functions never truly go to zero either and act to quantify potential discrepancies involved in modelling the physical system with deep networks, e.g. $\mathcal{L}_{f_{n_0}}$ can be understood as the outstanding error in approximating the neutral transport theory. Of these converged runs, the normalized mean loss functions at the end of training for the collection of realizations were found to be $\mathcal{L}_{GPI} = (1.44 \pm 0.42) \times 10^2$, $\mathcal{L}_{corr} = (-4.51 \pm 0.20) \times 10^{3}$, $\mathcal{L}_{relcorr} = 6.75 \pm 0.03$, and $\mathcal{L}_{f_{n_0}} = (3.57 \pm 3.79) \times 10^{-5}$. Values of the loss metrics remaining finite signify departures from exactly satisfying the training conditions being learnt. Identifying these errors allows for their iterative improvement, while identifying even better loss functions for generalized training is left for future work. 

 

\section{\label{sec:level5}Uncovering dynamics in experimental turbulence imaging in Alcator C-Mod}


\begin{figure*}[ht]
\includegraphics[width=1.0\linewidth]{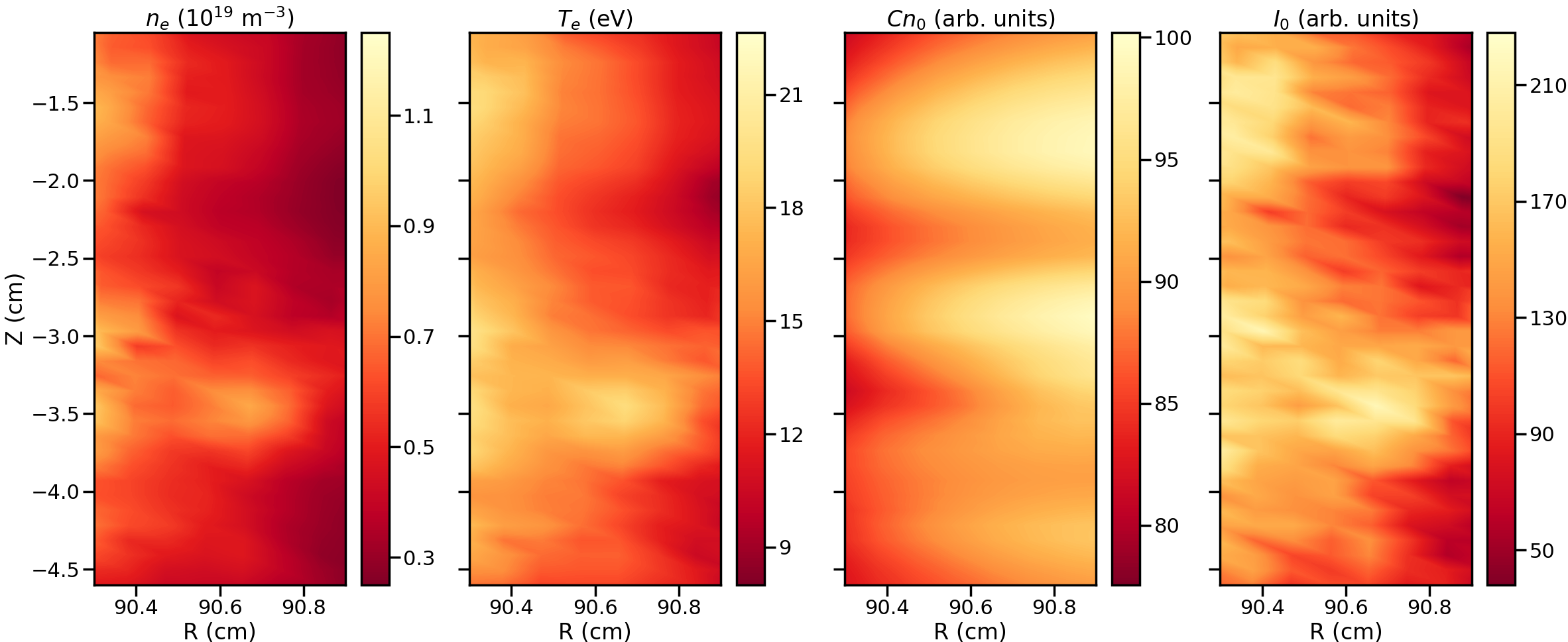}
\caption{\label{learned_2D_turbulence1}The learned 2-dimensional $n_e$, $T_e$, and $Cn_0$ for plasma discharge 1120711021 along with the experimentally observed 587.6 nm photon emission at $t = 1.312815$ s. The learned measurements are based upon the collective predictions within the deep learning framework training against the neutral transport physics and $N_P = 1$ CR theory constraints.}
\end{figure*}

\begin{figure*}[ht]
\includegraphics[width=1.0\linewidth]{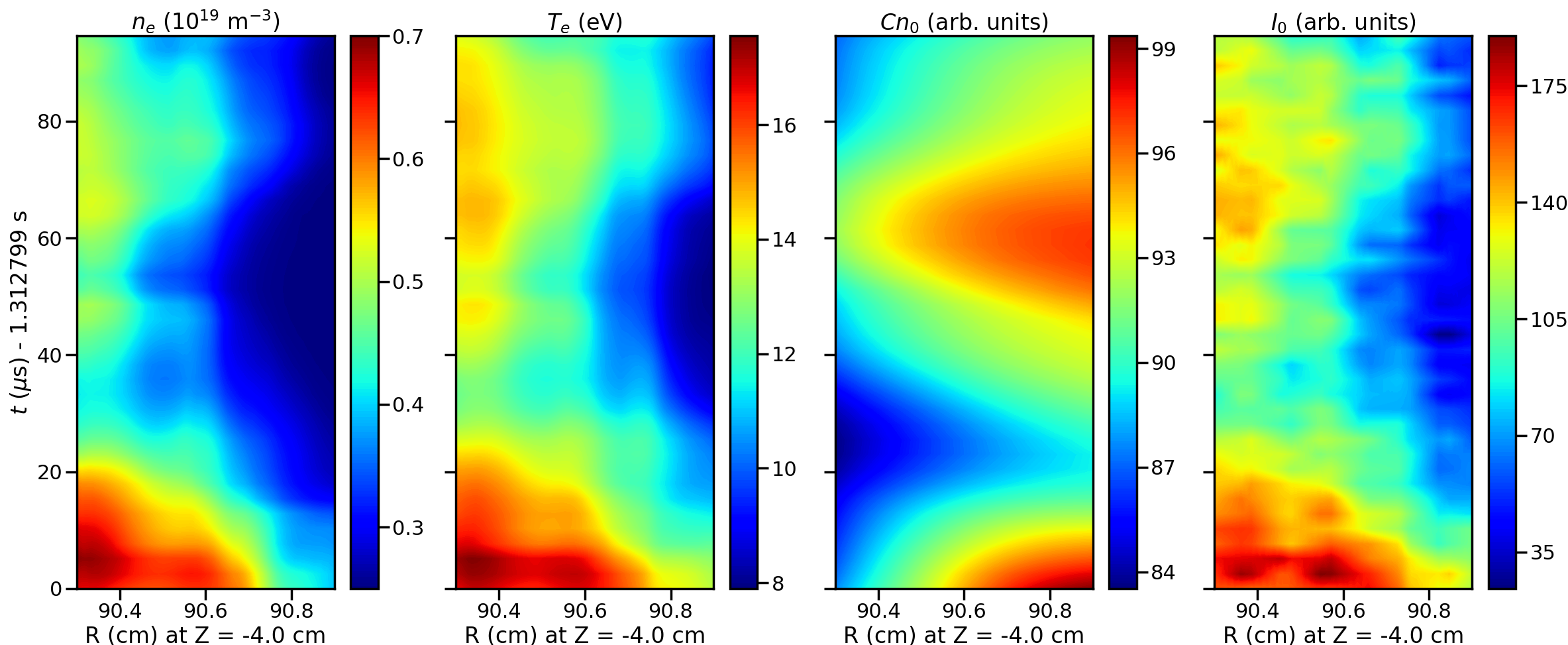}
\caption{\label{GPI_predvexp_1d_1120711021_rvt}The learned $n_e$, $T_e$, and $Cn_0$ along with the experimentally observed 587.6 nm photon brightness for plasma discharge 1120711021 at $Z = -4.0 \text{ cm}$. These quantities are plotted as a function of radius and time.}
\end{figure*}

The learned turbulent $n_e$, $T_e$, and $Cn_0$ from our time-dependent analysis of fast camera imaging for plasma discharge 1120711021 using an ensemble of 50 optimizers are visualized in 2-dimensional space along with experimentally observed GPI measurements in Figure \ref{learned_2D_turbulence1}. A correlation matrix for the normalized relative fluctuations in 2-dimensional space over the roughly 100 $\mu$s time window analyzed are displayed in Table \ref{table_correlation_matrix}.

\begin{table}[ht]
\centering
\renewcommand{\arraystretch}{1.}
\begin{NiceTabular}{ p{0.7cm}|p{1.1cm}|p{1.1cm}|p{1.1cm}|p{1.1cm}|>{\arraybackslash}p{1.1cm}| }[]
 {} & $\{n_e\}$ & $\{T_e\}$ & $\{n_0\}$ &$\{I^*\}$ &$\{I_0\}$\\ \cline{1-6}
 $\{n_e\}$ & 1.000 & 0.887 & -0.325 & 0.843 & 0.822
\\ \cline{1-6}
 $\{T_e\}$ & 0.887 & 1.000 & -0.307 & 0.925 & 0.902
\\ \cline{1-6}
 $\{n_0\}$ & -0.325 & -0.307 & 1.000 & -0.051 & -0.059
\\ \cline{1-6}
 $\{I^*\}$ & 0.843 & 0.925 & -0.051 & 1.000 & 0.971
\\ \cline{1-6}
 $\{I_0\}$ & 0.822 & 0.902 & -0.059 & 0.971 & 1.000
\\
\end{NiceTabular}
\caption{\label{table_correlation_matrix}A correlation matrix of the turbulent measurements inferred and observed experimentally in plasma discharge 1120711021. For reference, $I^*$ is the predicted emissivity given by Eq. \eqref{eq:emissivity_GPI}, and $I_0$ is the experimentally observed brightness of the 587.6 nm line. Each quantity's normalized fluctuation amplitude, i.e. $\{X\} = (X - \langle X \rangle)/\langle X \rangle$, is based upon measurements over $90.3 < R \ \text{(cm)} < 90.9$, $-4.6 < Z \ \text{(cm)} < -1.0$, and $1.312799 < t_{GPI} \ \text{(s)} < 1.312896$.}
\end{table}

\begin{figure*}[ht]
\includegraphics[width=1.0\linewidth]{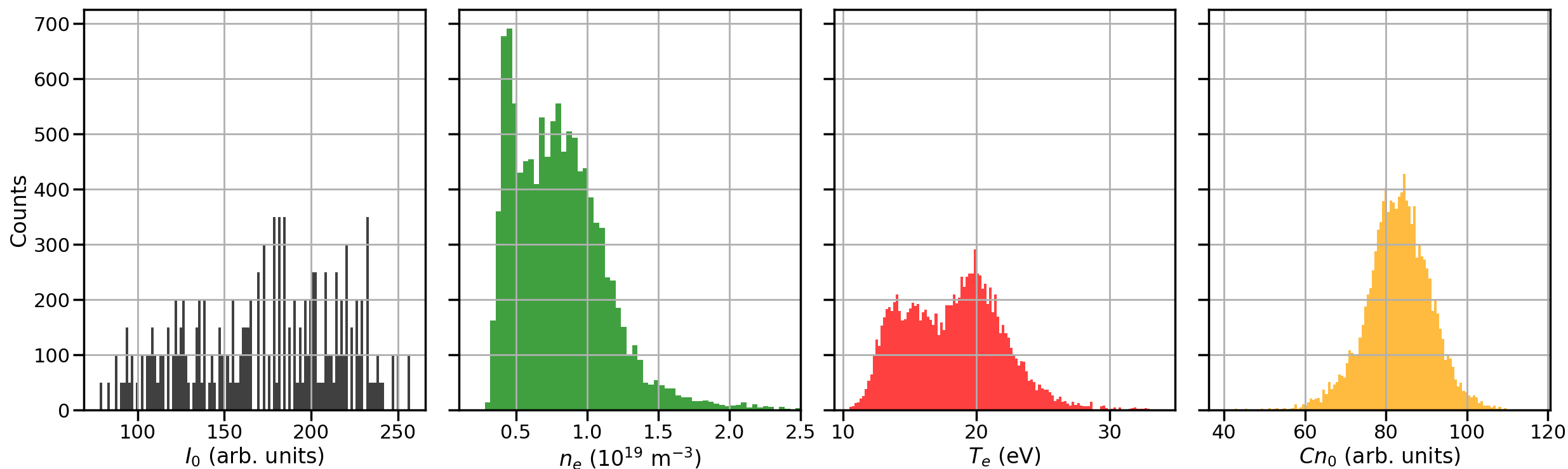}
\caption{\label{turbulence_histogram}Histograms displaying the distribution of turbulent $n_e$, $T_e$, and $Cn_0$ at [$R$ = 90.3 cm, $-4.6 < Z \ \text{(cm)} < -1.0$, $1.312799 < t_{GPI} \ \text{(s)} < 1.312896$] along with the experimentally observed 587.6 nm line intensity.}
\end{figure*}

The positive fluctuations in brightness are largely correlated with $n_e$ and $T_e$, and these regions tend to have depressed values of $n_0$ as the ionization rate is elevated. This results in a ``shadowing effect" in atomic helium trajectories arising from fluctuation-induced ionization. The autocorrelation time of $n_0$ also decreases with radius, while it increases for $n_e$ and $T_e$. Temporal variation with radius is visualized in Figure \ref{GPI_predvexp_1d_1120711021_rvt} where, considering a 1-dimensional slice of Figure \ref{learned_2D_turbulence1}, the same physical quantities are plotted at $Z = -4.0$ cm. While correlations vary poloidally and radially, and precise dependencies across the turbulent variables change as $n_e$ and $T_e$ increase, the observed line emission is found to be strongly correlated with electron pressure. The atomic helium density fluctuations do not vary directly proportional to $I_0$ in this far edge region on open field lines near the gas tubes. There is instead a weak negative correlation over this narrow radial extent arising from the largest brightness fluctuations corresponding to trajectories with elevated ionization rates causing a depletion, or shadowing, of HeI. The maximal $n_0$ fluctuation amplitudes tend to be roughly 30--40\% from peak-to-trough in this far edge region which sits away from the LCFS, where sharper equilibrium gradients and smaller relative fluctuation levels may result in different correlations. And while relative fluctuations may be correlated from $90.3 < R \ \text{(cm)} < 90.9$ as in Table \ref{table_correlation_matrix}, connections between the turbulent quantities are nonlinear. To better visualize their interdependence, Figure \ref{turbulence_histogram} displays histograms for $n_e$, $T_e$, and $Cn_0$ vertically along $R = 90.3$ cm (with $\psi_n$ in the range 1.035 -- 1.053). The fluctuations follow different statistical distributions and cannot necessarily be linearly mapped from the noisy HeI line intensity measured by the fast camera.



\begin{figure*}[ht]
\includegraphics[width=1.0\linewidth]{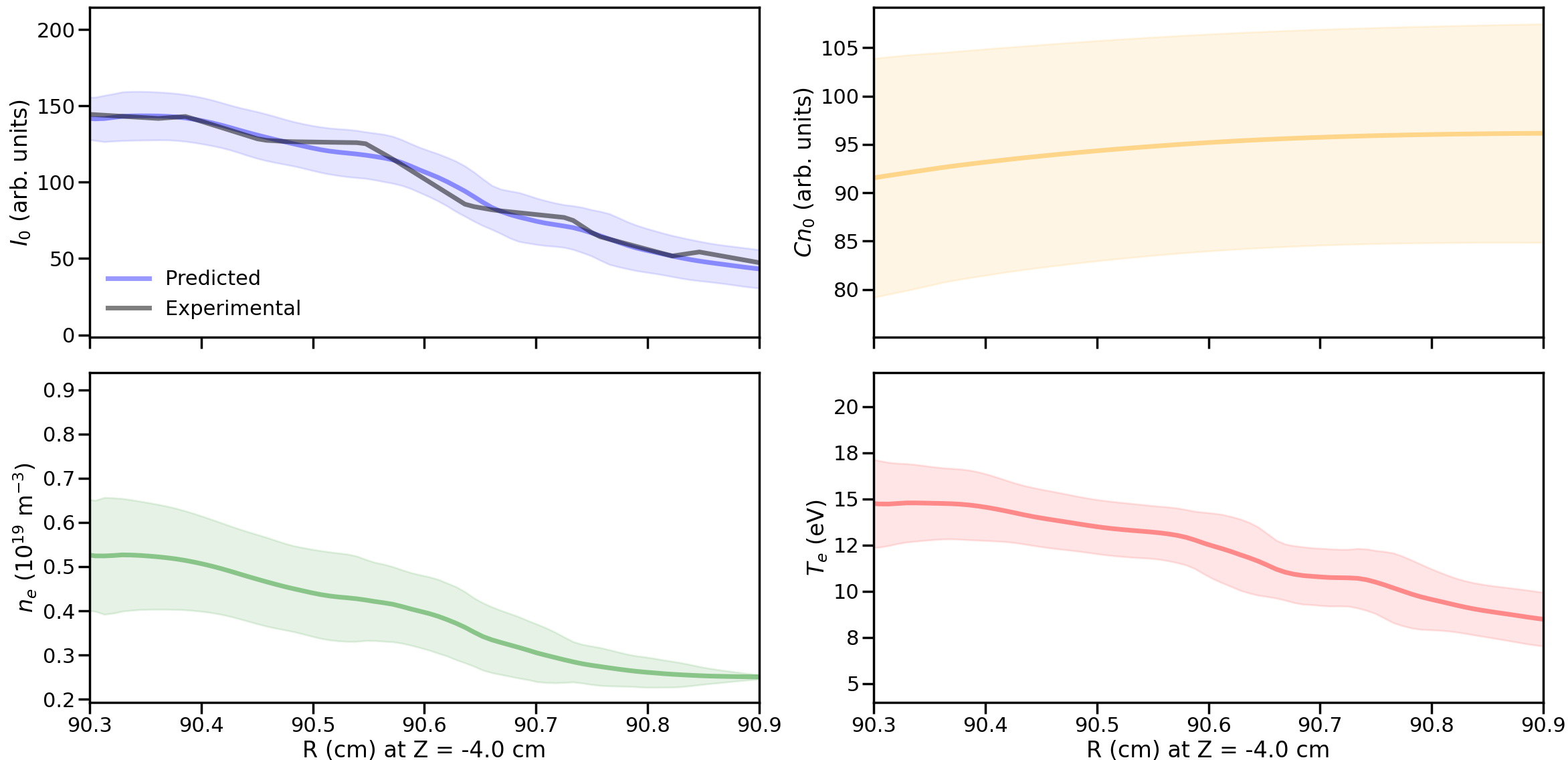}
\caption{\label{predicted_1D}Radial profiles of the inferred turbulent $n_e$, $T_e$, and $Cn_0$ at $[Z = -4.0 \text{ cm}, t = 1.312866 \text{ s}]$ along with a trace of the experimentally observed and predicted GPI intensity profiles. The computed line emission is based upon the deep learning framework following Eq. \eqref{eq:emissivity_GPI}. The dark line in each plot corresponds to the average output of the ensemble of realizations, while the shaded uncertainty intervals correspond to scatter ($\pm 2\sigma$) arising from the independently trained networks.}
\end{figure*}


\begin{figure}[ht]
\includegraphics[width=1.0\linewidth]{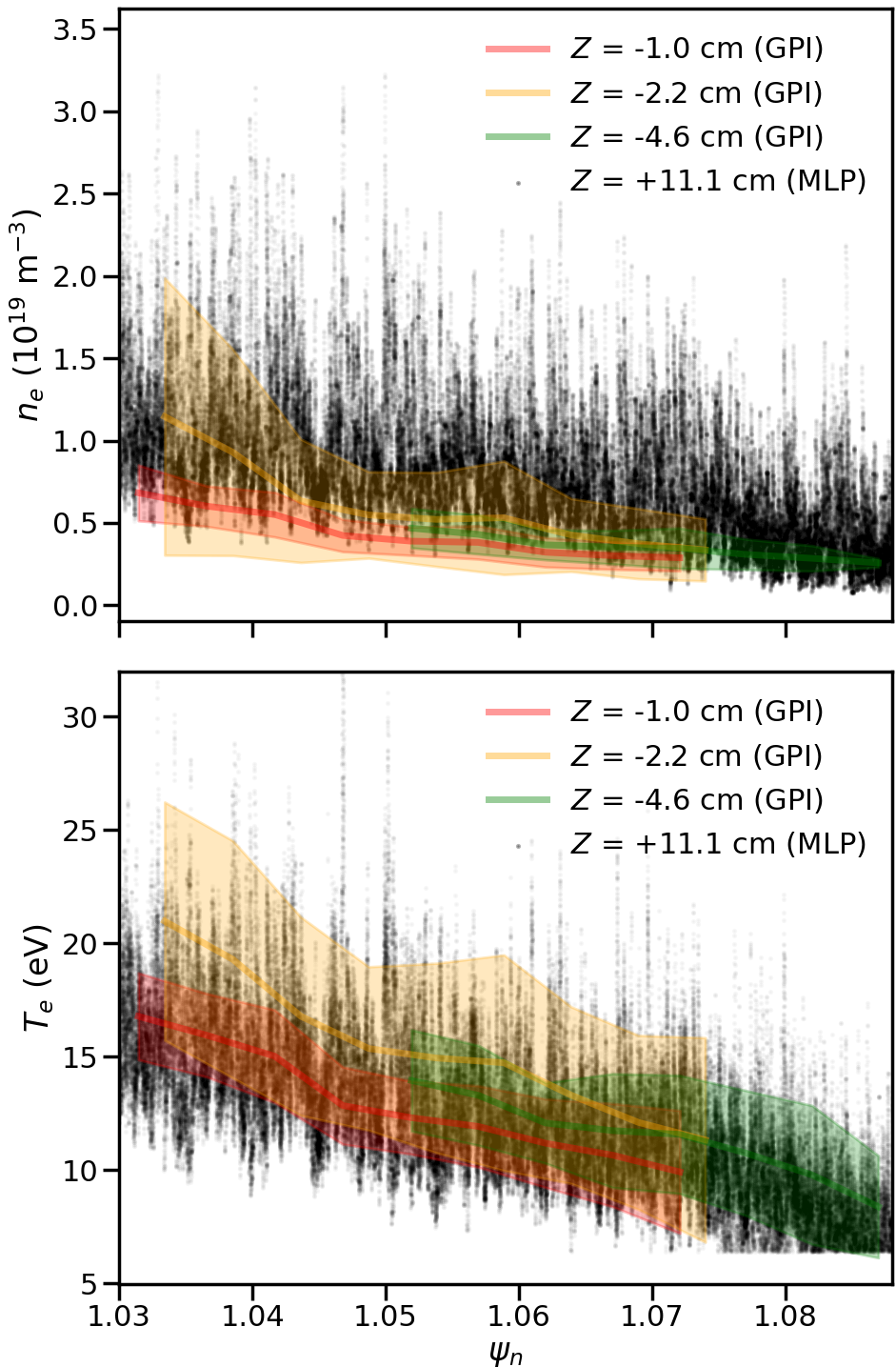}
\caption{\label{GPIvprobes}For evaluation of radial profiles, we compute the time-averaged $n_e$ and $T_e$ over roughly 100 $\mu$s from the experimental GPI at different vertical locations. These quantities inferred from GPI are compared against independent scanning mirror Langmuir probe measurements from 4 electrodes which are not time-averaged. This MLP is located at $Z = +11.1$ cm with a radially moving probe head in the edge plasma. Shaded intervals on the time-averaged $n_e$ and $T_e$ inferred from GPI correspond to the temporal scatter ($\pm 2\sigma$) associated with the learned turbulent fluctuations in our optimization framework. All measurements are mapped to normalized poloidal magnetic flux coordinates, $\psi_n$.}
\end{figure}


We note that these learned $n_e$, $T_e$, and $Cn_0$ are consistent solutions with the collisional radiative and optimization constraints being trained upon, but not necessarily unique solutions. Accordingly, in Figure \ref{predicted_1D}, we display the predicted light emission from the ensemble of realizations against the fast camera's measurements. We also plot the mean outputs and uncertainty intervals for the turbulent $n_e$, $T_e$, and $Cn_0$ associated with the scatter of running an ensemble of our stochastic realizations. There is no temporal averaging of the profiles in Figure \ref{predicted_1D}. For GPI on Alcator C-Mod, sharp features exist in the experimental data potentially associated with noise, while the learned line intensity from the collection of networks is smoother yet still largely consistent with the observed brightness. These measurements enable novel research pathways into the 2-dimensional tracking of experimental parameters (e.g. turbulent particle and heat fluxes both radially and poloidally) and calculation of fluctuating fields for turbulence model validation \cite{Mathews2021PRE,Mathews2021PoP}. They further provide the first quantitative estimates of 2-dimensional structure for neutrals in an experimental fusion plasma on turbulent scales. 


To further examine the validity of these results, we juxtapose turbulent $n_e$ and $T_e$ from our GPI measurements of the single spectral line against an independent MLP with 4 electrodes. This scanning probe is plunged at times overlapping with our gas puff analysis, i.e. $1.287631 < t_{MLP} \ \text{(s)} < 1.317671$ versus $1.312799 < t_{GPI} \ \text{(s)} < 1.312896$, although located at different positions toroidally and vertically. The MLP measures fluctuations in time as it scans through the edge plasma, and we consider its resultant constructed radial profile. For the purpose of comparison, we time-average our turbulent measurements at different $Z$-locations to compare the $n_e$ and $T_e$ fluctuations. (Comparison of $n_e$ and $T_e$ without time-averaging can be found in \cite{Mathews_thesis}.) While the measurement regions spatially spanned by the two independent diagnostics are magnetically disconnected, we map the fluctuations to common poloidal magnetic flux coordinates based upon magnetohydrodynamic equilibrium reconstruction in the tokamak plasma using the EFIT code \cite{Lao_1985}. Deconstructing the GPI fluctuations into the turbulent $n_e$, $T_e$, and $n_0$ instead of the raw brightness from atomic emission largely resolves diagnostic misalignment challenges \cite{LaBombard_probevGPI} and, in contrast with past analysis of plasma discharge 1120711021 \cite{Russell_probevGPI}, there is no radial shift applied to align the turbulent fluctuation profiles from these two independent experimental diagnostics. Profiles on these open field lines fluctuate strongly with perturbations of up to 10 - 100\%. For turbulence measurements by GPI in this 2-dimensional spatial domain spanning approximately $100 \ \mu$s, peak $n_e$ and $T_e$ fluctuations do not far exceed $2.5 \times 10^{19} \ \text{m}^{-3}$ and 25 eV, respectively, which are roughly consistent with the MLP measurements in the far SOL. When evaluating the two sets of measurements side-by-side in Figure \ref{GPIvprobes}, we find excellent agreement in magnitude and structure between the $T_e$ measurements. For the electron density channel, the MLP $n_e$ data are slightly elevated on average although still quantitatively consistent within the measurement bounds of the 4 electrodes. A potential contributing factor to this observed difference in $n_e$ peaks could be natural variations in the poloidal structure of tokamak turbulence over the narrow time window, although the inferred density fluctuations are still large (on the order of 10--100\%). It is worth noting that the MLP scan in the plot lasts roughly 6000 $\mu$s (i.e. 60$\times$ longer than the duration of the GPI analysis). Temporal averaging over longer durations can help reduce the imprint of intermittent plasma transport when comparing radial profiles at different vertical positions \cite{Mathews_thesis}. One should also remember that, beyond the diagnostics viewing different spatiotemporal locations, systematic uncertainties extant in both the GPI and probe measurements can cause discrepancies left to be reconciled \cite{hutchinson_2002,probe_review}. For example, the MLP is intrinsically perturbative to local edge plasma conditions and experimental analysis of the probe edge sheath assumes electrons can be described by a single Maxwellian velocity distribution \cite{kuang_thesis}. Additionally, while our training paradigm attempts to find consistent solutions within the applied optimization framework, questions of uniqueness and generalized constraints are still being explored.







\section{\label{sec:level6}Conclusion}

In summary, we have developed a novel time-dependent deep learning framework for uncovering the turbulent fluctuations of both the plasma quantities, $n_e$ and $T_e$, as well as the neutrals underlying experimental imaging of HeI line radiation. Significantly, this allows determination of 2-dimensional fluctuation maps in the plasma boundary, revealing detailed spatiotemporal structure and nonlinear dynamics. It thereby extends the usefulness of the gas puff imaging method. The computational technique intrinsically constrains solutions via collisional radiative theory and trains networks against neutral transport physics. This advancement has allowed for the first estimates of the 2-dimensional turbulent $n_e$, $T_e$, and $n_0$ which reveal fluctuation-induced ionization effects in a fusion plasma based upon optical imaging of just the 587.6 nm line. While our analysis is demonstrated on the edge of the Alcator C-Mod tokamak with quantitative agreement found with independent probe measurements, this technique is generalizable to ionized gases in a wide variety of conditions and geometries (e.g. stellarators, spheromaks, magneto-inertial fusion).

We emphasize that this is just the beginning of this technique and that a number of opportunities for future development exist both computationally and experimentally. One key outstanding question is the identification of underlying numerical and physical factors contributing to non-uniqueness in outputs during optimization. From experimental noise to the chaotic properties of the turbulent system, finding sufficient conditions for precise convergence is the focus of ongoing research. Future extensions of the framework also include expanding the radial domain of coverage towards closed flux surfaces, which will require widening the queried bounds on $n_e$, $T_e$, $n_0$, and improving the overall training paradigm via adaptive training and architecture structures \cite{wang2020understanding}. For example, neutral density amplitudes can vary over orders of magnitude with steep shapes in background equilibrium profiles. Tactfully embedding this information during training of the networks can aid with the overall physical modelling via optimization. In this way, better experimental constraints from 1-dimensional data may help uncover further dynamics not otherwise directly probed. 

Adaptation to other experiments is a logical next step, and translating our present technique to contemporary experimental devices using helium beams is a pathway that can be explored immediately for regions that are traditionally difficult to probe (e.g. X-point regions). Alternatively, our deep learning framework does not apply any discretization of the spatiotemporal domain---networks provide natural continuum representations for dynamical variables---and can thus be extended in principle to 3-dimensional geometries as well to account for integrated light emission along the camera's lines-of-sight if using wide gas distributions as expected, for example, when running He plasmas on ITER \cite{ITER_2007}. Further, this global turbulence imaging technique provides new ways to diagnose high pressure plasma events, e.g. disruptive instabilities such as edge localized modes that can be destructive to plasma facing components. Translating our framework for direct analysis of deuterium instead of helium is also possible with a few modifications, but requires investigation of relevant CR physics \cite{Greenland_full} where charge exchange and molecular effects are no longer necessarily negligible \cite{AMJUEL}. One prospect is to couple the turbulent $n_e$ and $T_e$ learned within our framework with Monte Carlo neutral transport codes \cite{STOTLER_time_dependent_DEGAS2}, potentially allowing recovery of 2-dimensional time-dependent estimates of atomic and molecular deuterium density and its emissivity, e.g. through the ultraviolet Ly$_\alpha$ line. These could be compared directly to experimental measurements of line emission from deuterium \cite{Lyalpha0,Lyalpha1}. Such extended comparisons with neutral dynamics will be important in the testing of reduced edge plasma turbulence models \cite{Mathews2021PRE}.





 

\begin{acknowledgments}
\noindent We wish to thank  M. Francisquez, F. Sciortino, A. Thrys\o{}e, and T. Greenland for insights shared and helpful discussions; P. Maybank and L. Mulholland from the Numerical Algorithms Group (NAG) for technical support; D. Brunner for operation of the probe. All simulations presented and codes run were performed using MIT's Engaging cluster and we are grateful for the team's assistance with computing resources. The work is supported by the Natural Sciences and Engineering Research Council of Canada (NSERC) by the doctoral postgraduate scholarship (PGS D), Manson Benedict Fellowship, and the U.S. Department of Energy (DOE) Office of Science under the Fusion Energy Sciences program by contracts DE-SC0014264, DE-SC0014251, and DE-AC02 09CH11466. Relevant data and supplementary files are available from the corresponding author.
\end{acknowledgments}

\bibliography{main.bib}
\end{document}
%